\documentclass[10pt,onecolumn]{IEEEtran}
\input{epsf}
\usepackage{framed}
\usepackage{color}

\definecolor{bgrd}{rgb}{1,1,1}
\definecolor{grey}{rgb}{0.9,0.9,0.6}
\definecolor{gray}{rgb}{0.5,0.5,0.5}
\definecolor{dkr}{rgb}{0.6,0.2,0.2}
\definecolor{dkg}{rgb}{0,0.5,0}
\definecolor{dkb}{rgb}{0.0,0.1,0.7}
\definecolor{light-gray}{gray}{0.85}

\newcommand{\test}{\mbox{$\begin{array}{c}
\stackrel{\stackrel{\textstyle {\cal H}_1}{\textstyle >}}
{\stackrel{\textstyle <}{\textstyle {\cal H}_0}} \end{array}$}}

\usepackage{theorem}
\newtheorem{theorem}{Theorem}

\newtheorem{corollary}{Corollary}

\renewcommand{\P}{\mathbb{P}}

\newcommand{\beq}{\begin{equation}}
\newcommand{\eeq}{\end{equation}}
\newcommand{\beqa}{\begin{eqnarray}}
\newcommand{\eeqa}{\end{eqnarray}}
\newcommand{\dfz}{\triangleq}

\newcommand{\bq}{\mbox{\boldmath{$q$}}}

\newcommand{\bA}{\mbox{\boldmath{$A$}}}

\newcommand{\bd}{{\mathbf{d}}}

\newcommand{\bu}{{\mathbf{u}}}
\newcommand{\bz}{{\mathbf{z}}}

\newcommand{\bm}{{\mathbf{m}}}

\newcommand{\bt}{{\mathbf{t}}}
\newcommand{\bw}{{\mathbf{w}}}
\newcommand{\bv}{{\mathbf{v}}}
\newcommand{\ba}{{\mathbf{a}}}
\newcommand{\bh}{{\mathbf{h}}}

\newcommand{\bx}{{\mathbf{x}}}

\newcommand{\VAR}{\textnormal{$\mathbb{V}$}}
\newcommand{\V}{\mathbb{V}}

\newcommand{\E}{\mathbb{E}}

\newcommand{\cE}{{\cal E}}

\newcommand{\cI}{{\cal I}}

\newcommand{\cN}{{\cal N}}

\newcommand{\bcE}{\bm{\mathcal{E}}}

\newcommand{\cH}{{\cal H}}
\newcommand{\cA}{{\cal A}}

\usepackage{cite}
\usepackage{multirow}
\usepackage{rotating}
\usepackage{hhline}
\usepackage{amssymb}
\usepackage{mathrsfs}
\usepackage{graphicx}
\usepackage{epstopdf}
\usepackage{amsmath}
\usepackage{empheq}

\usepackage{bm}

\usepackage{soul}
\usepackage[vlined,ruled]{algorithm2e}

\usepackage{xcolor, framed}

\begin{document}

\title{Decision Learning and Adaptation \\ over Multi-Task Networks} 

\author{Stefano~Marano and Ali H. Sayed  
\thanks{S.~Marano is with DIEM, University of Salerno, via Giovanni Paolo~II 132, I-84084, Fisciano (SA), Italy (e-mail: marano@unisa.it).} 
\thanks{A.~H.~Sayed is with the Ecole Polytechnique Federale de Lausanne EPFL, School of Engineering, CH-1015 Lausanne, Switzerland (e-mail: ali.sayed@epfl.ch).}
} 

\maketitle

\begin{abstract}
This paper studies the operation of multi-agent networks engaged in multi-task decision problems under the  paradigm of simultaneous learning and adaptation. 
Two scenarios are considered: one in which a decision must be taken among multiple states of nature that are known but can vary over time and space, and another in which there exists a known ``normal'' state of nature 
and the task is to detect unpredictable and unknown deviations from it. 
In both cases the network learns from the past and adapts to changes in real time in a multi-task scenario with different clusters of agents addressing different decision problems. The system design takes care of challenging situations with clusters of complicated structure, and the performance assessment is conducted by computer simulations.
A theoretical analysis is developed to obtain a statistical characterization of the agents' status
at steady-state, under the simplifying assumption that clustering is made without errors. 
This provides approximate bounds for the steady-state decision performance of the agents. Insights are provided for deriving accurate
performance prediction by exploiting the derived theoretical results.

\end{abstract}

\begin{IEEEkeywords}
Learning and adaptation, distributed detection, multi-task networks, diffusion schemes, ATC rule.
\end{IEEEkeywords}

\section{Introduction}
\label{sec:intro}

\IEEEPARstart{L}{earning} and adaptation in multi-agent decision systems impose related but contrasting requirements. 
This is because learning deals with the ability of a network to learn from
agents' observations in order to deliver accurate decisions about a phenomenon of interest (state of nature), while adaptation deals with 
the ability of tracking state changes over time and monitoring statistical drifts.
Learning is enhanced by maintaining a long memory of the past, while adaptation requires to assign larger weights to recent observations and progressively neglect older
measurements. Thus, the design of learning and adaptation systems requires a careful tradeoff between opposite needs.

This paper focuses on \emph{multi-task} networks in which agents are grouped into clusters 
characterized by different observation models. These networks generalize the operations of their single-task counterpart and are relevant when the phenomenon of interest is space-dependent, yielding inhomogeneous observations at agents located in different regions of the surveyed area. The main challenge is that agents are not aware of which cluster they belong to, and fusing information from their neighbors without awareness of the multi-task structure can be self-defeating. It is not obvious how to take advantage of the network structure, and a careful data diffusion mechanism must be designed.

This paper addresses the design of a diffusion mechanism for two kinds of decision problems and analyzes the resulting decision performance.
The first decision problem is a multihypothesis test with known but time/space varying states of nature. The second problem is a binary test
in which there is a known ``normal'' state of nature and the task is to detect unknown time/space varying deviations thereof.

\subsection{Related Work}

The literature addressing distributed detection and inference problems by means of networked agents is abundant, see e.g.,\cite{chamberland,PreddKulkarniPoor-magazine06,akyildiz-survey,Varshney:book,viswanathan97,blum97,tsitsiklis93,Tsitsiklis88} and the references therein. In a star-topology architecture, agents deliver data to a central unit to where the inferential task is processed. The central unit can be located in a fixed position or can travel across the surveyed area to facilitate communication with agents~\cite{SENMA,doasplet05,spawc06,tong:C-SENMA-LDPC}. Substantial advantages in terms of tolerance to failures, security, and robustness
are obtained in fully-flat architectures not equipped with a central unit, in which the inference is obtained by distributed processing and local communications among nearby agents.
Examples of these implementations can be found in~\cite{boyd-infocom,running-cons,asymptotic-rc,kar-moura-stsp,mouraetal2011,mouraetal2012,mouraetal2012_2} 
for consensus strategies and in~\cite{CattivelliSayedEstimation,CattivelliSayedDetection,ZhaoSayedLMSestimation,TuSayedConsensus,SayedSPmag,SayedNOW2014,SayedprocIEEE,chen-sayed-IT1,chen-sayed-IT2,MattaSIPN16,TowficChenSayedIT2016,BracaetalIT,MaranoSayedIT19} for diffusion strategies. 
In more recent works, the original formulation of single-task networks has been extended to address multi-task scenarios, 
e.g.,~\cite{ChenJSTSP17,KhawatmiSP17,NassifSP17,NassifSPL19,KhawatmiSP2019}. These works focus mainly on
estimation problems, and our main contribution is to provide generalizations for \emph{decision} problems, which have not received enough attention so far over multi-task adaptive networks.
A preliminary version of one of the decision problems addressed here is discussed on the conference article~\cite{ICASSP2020submitted}.


\subsection{Contribution and Organization}
This article addresses two multi-task decision problems by designing cluster-aware diffusion mechanisms.
Section~\ref{sec:genesis} describes the genesis of the LMS (least-mean-square) algorithm for single-agent decision making, with emphasis on how the algorithm is adapted from the estimation to the decision context. 
The role of this introductory part is twofold. First, it provides a motivation for the diffusion algorithms introduced later and, second, sheds some light on possible 
extensions, which are however left for future studies. 
In Sec.~\ref{sec:ATC} the diffusion algorithm for multi-agent single-task networks is presented. This algorithm, called ATC (adapt-then-combine) diffusion, is the building block for the decision procedures
over multi-task networks studied in the paper.
The design of multi-task decision systems exploits suitable adaptations of the ATC algorithm and is presented in Secs.~\ref{sec:IA} and~\ref{sec:PIA}, where two different decision problems over multi-task networks are posed and addressed. For both problems, a theoretical analysis leading to the statistical characterization of the agents' status at steady-state is conducted in Sec.~\ref{sec:approx}. The results of computer simulations are discussed in Sec.~\ref{sec:comp}. In addition, in Sec.~\ref{sec:comp}, the statistical characterization developed in Sec.~\ref{sec:approx} is exploited to derive approximate bounds on the error probability of the decision systems. Section~\ref{sec:concl} contains final remarks.

\subsection{Notation}

Boldface symbols denote random variables and normal font their realizations and deterministic quantities.
Matrices are shown in (non-caligraphic) capital letters, while small letters are reserved to both scalar and column vectors, with the exception of the scalar integers $M$, $S$, $H$, $R$,
and $N$.
For scalar quantities, the time index (or algorithm iteration number) is enclosed in parentheses, while the agent label is shown as a subscript. 
Thus, for instance, $\bx_k(n)$ denotes the random scalar $\bx$ at time~$n$ referring to agent~$k$. 
Conversely, in the case of vectors, the time dependence is indicated by a subscript, as, for example, $\bu_i$ denotes a random vector $\bu$ evaluated at time $i$. 
Superscript $T$ denotes vector transposition. 
Statistical expectation, variance, and probability operators are denoted by $\E$, $\VAR$, and $\P$, respectively. 
They always are computed under the hypothesis in force, and a subscript is added whenever it is appropriate to emphasize this fact.
When a-priori probabilities are assigned to these hypotheses --- the Bayesian setting --- the subscript indicates the conditioning. 
The probability that a random variable takes value $a \in \cA$, where $\cA$ is a finite alphabet, under the probability model $\cH_h$, is denoted by $p_h(a)$. The corresponding probability mass function (PMF) is denoted by the row vector $p_h$. In the Bayesian setting, these PMFs are conditioned to the hypothesis.

\section{Single Agent: Genesis of the LMS Algorithm for Decision}
\label{sec:genesis}

One key ingredient of the network diffusion algorithms studied in this article is the least-mean-square (LMS) algorithm for decision,
which is introduced in this section in connection with a multi-hypothesis decision problem involving a single agent. 
We first consider in Sec.~\ref{sec:SALMSest} an estimation problem, which is the context in which the LMS algorithm is usually developed. In Sec.~\ref{sec:SALMSdec}, the version of the algorithm tailored to decision problems is presented.
Some known facts about multi-hypothesis decisions are recalled in Sec.~\ref{sec:multi}, which are then used as guideline for the algorithm design of Sec.~\ref{sec:LMSmulti}.

\subsection{LMS Algorithm for Estimation}
\label{sec:SALMSest}

Let us start by considering an estimation problem with a single agent. Let $\bd\in \Re$ be a zero-mean scalar random variable with variance $\E\bd^2>0$, and $\bu \in \Re^{M}$ a zero-mean random vector with positive-definite covariance matrix $\E \bu \bu^T >0$. The quantity $\bd$ is unknown while $\bu$ is observed.
The goal is to solve the optimization problem $\min_w J(w)$,
where $w \in \Re^{M}$ is a weight vector, and $J(w) : \Re^{M} \mapsto \Re$ represents a cost function that quantifies the penalty incurred when the unknown $\bd$ 
is replaced by the linear transformation $\bu^T w$ of the observation.
One common choice is the quadratic cost function $J(w)=\E(\bd -\bu^T w)^2$, in which case the solution $w^{\rm o}$ 
is given by $w^{\rm o}=(\E \bu \bu^T)^{-1} \E \bd\bu$,
and the \emph{linear least-mean-square estimator} of $\bd$ given $\bu$ is $\widehat \bd=\bu^T w^{\rm o}$~\cite[Th.~8.1, p.~142]{Sayed2008adaptive}.

A recursive solution to the optimization problem $\min_w J(w)$ 
with quadratic cost function is provided by the \emph{steepest-descendent} algorithm: set $w_{0}$ equal to some initialization vector, and iterate as follows:
\begin{align}
w_i=w_{i-1} + \mu \big [\E \bd\bu- \E \bu \bu^T \,  w_{i-1} \big ], \quad i=1,2,\dots,
\label{eq:steepest}
\end{align}
where the step-size $\mu>0$ is sufficiently small (less than 2 divided by the largest eigenvalue of matrix $\E \bu \bu^T$), see~\cite[Th.~8.2, p.~147]{Sayed2008adaptive}.
It can be shown that $\E \bd\bu- \E \bu \bu^T \,  w_{i-1} =-\nabla J(w_{i-1})$, which makes it possible to rewrite~(\ref{eq:steepest}) in terms of the gradient vector $\nabla J(w_{i-1})$. The resulting expression is useful when alternative cost functions are used.

What is especially relevant in the adaptive framework is the consideration that the quantities $\E \bu^T \bu$ and $\E \bd\bu$ may not be known and are expected to vary over time. 
In these situations, assuming that we have access to streaming data in the form of a sequence of realizations $\{d(i), u_i\}_{i\ge 1}$ of $\bd$ and $\bu$, a viable alternative to~(\ref{eq:steepest}) is obtained if we drop the expectation signs and replace the random variables by their current realizations, yielding the following algorithm: set $w_{0}=$ some initial guess, 
\begin{align}
w_i=w_{i-1} + \mu u_i \big [d(i)-  u_i^T \,  w_{i-1} \big ], \quad i=1,2,\dots,
\label{eq:LMS0}
\end{align}
with a sufficiently small $\mu$.
This \emph{stochastic gradient} approximation (because the true gradient is replaced by a \emph{noisy} version thereof) is known as the LMS algorithm, see~\cite[Th.~10.1, p.~166]{Sayed2008adaptive}.
The LMS algorithm learns the data statistics and at the same time is able to track statistical drifts, which are essential characteristics for the design of cognitive intelligent inference systems with learning and adaptation properties. 

\subsection{LMS Algorithm for Decision}
\label{sec:SALMSdec}

Suppose $M=1$, namely ${w_i}={w(i)}$ and ${u_i}={u(i)}$ are scalars, and suppose also $u(i)=1$ for all $i$. Formal substitution in~(\ref{eq:LMS0})
gives: $w(0)=0$, 
\begin{align}
w(i)&=w(i-1)+ \mu  [ d(i)- w(i-1)], \qquad i\ge1. \label{eq:LMSdet}
\end{align}
In this article we focus on the version of the algorithm shown in~(\ref{eq:LMSdet}) that will be referred to as LMS ``for decision''.
Note that the the right-hand side of~(\ref{eq:LMSdet}) is a convex combination: $\mu d(i) + (1-\mu) w(i-1)$.

By assuming independent and identically distributed (IID) data $\{\bd(i)\}_{i\ge1}$, and iterating~(\ref{eq:LMSdet}), we get the output of the LMS algorithm for decision in the form:
\begin{align}
\bw(i) = \sum_{k=0}^{i-1} \mu (1-\mu)^k \bd(i-k),
\end{align}
and we have
\begin{subequations}
\begin{align}
\E  \bw(i) &=  \, [1-(1-\mu)^i] \, \E \bd,  \\
\VAR \bw(i) & = [1-(1-\mu)^{2i}] \frac{\mu}{2-\mu} \, \VAR \bd.
\end{align}
\label{eq:EVAR}%
\end{subequations}
From~(\ref{eq:EVAR}), we see that the output of the algorithm approximates $\E \bd$ when the number $i$ of iterations is sufficiently large and 
the step-size $\mu$ is $\ll 1$. This property, along with the inherent adaptation ability, motivates the use of~(\ref{eq:LMSdet}) in decision problems, as will become evident in the following.

\subsection{Multi-Hypothesis Decision and its Geometry}
\label{sec:multi}
Consider a decision problem involving $H$ exhaustive and mutually exclusive hypotheses $\cH_1,\dots,\cH_H$, and suppose the observed data $\{ \bx(i)\}_{i=1}^n$
are all drawn from only one of these distributions. We model the data as a random process made of conditionally IID random variables, where the conditioning is with respect to the true hypothesis. 
Henceforth, we focus on the case that the data are drawn from a finite alphabet~$\cA$, common to all the probability models $\cH_1,\dots,\cH_H$. 
The finite-alphabet setting is especially relevant for modern applications. 
In addition, with finite alphabets, the geometrical interpretation of the multi-hypothesis test is particularly simple and instructive, as we shall see soon.
Actually, most results of Sec.~\ref{sec:IA} can be easily generalized to include continuous observations, but the approach pursued in Sec.~\ref{sec:PIA} is founded on the empirical distribution of the observations, and the generalization to the continuous case is less obvious. 

In multi-hypothesis testing with uniform priors (all the hypotheses have the same occurrence probability) and conditionally IID data $\{ \bx(i)\}_{i=1}^n$, it is well known that the maximum likelihood (ML) decision
$\widehat \bh$ minimizes the error probability $P(e)$, where~\cite{poorbook}: 
\begin{subequations} 
\begin{align}
&\widehat \bh =\arg\max_h \frac 1 n \sum_{i=1}^n \log p_h(\bx(i)), \label{eq:hath} \\
&P(e)=\frac 1 H \sum_{h=1}^H \P_h(\widehat \bh \neq h). \label{eq:pe}
\end{align}
\label{eq:optcase1}%
\end{subequations}
In~(\ref{eq:hath}), $p_h(\bx(i))$ is the conditional PMF of $\bx(i)$ given $\cH_h$.
To avoid trivialities, we assume that these PMFs are strictly positive, 
$p_h(a)>0$, $\forall a \in \cA$, and all distinct, namely no two of these can be equal at all points $a\in\cA$. 
For simplicity, we also exclude the possibility of multiple maxima in~(\ref{eq:hath}) that would require randomized decisions.

Expression~(\ref{eq:hath}) can be rewritten in the following form~\cite[pp. 377-378]{CT2}:
\begin{align}
\arg\max_h \frac 1 n \sum_{i=1}^n \log p_h(\bx(i)) = \arg\min_h D(t_{\bx(1:n)} || p_h),
\label{eq:argmaxargmin}
\end{align}
where 
\begin{align}
D(p||q)=\sum_{a\in\cA} p(a) \log \frac{p(a)}{q(a)}
\end{align} 
denotes Kullback-Leibler (KL) distance~\cite{CT2} between PMFs $p$ and $q$, and 
$t_{\bx(1:n)}$ is the empirical PMF of the sequence $\bx({1:n})=[\bx(1), \bx(2),\dots,\bx(n)]$, defined as 
\begin{align}
\hspace*{-6pt} t_{\bx(1:n)}(a) = \frac{\textnormal{\small No. of occurrences of $a$ in $\bx({1:n})$}}{n}, \quad a \in \cA .
\end{align} 
An insightful geometrical interpretation of~(\ref{eq:optcase1}) immediately
follows~\cite{Westover}. To minimize $P(e)$ in~(\ref{eq:pe}): $(i)$~locate the type (or empirical PMF) of the observations in the probability simplex; $(ii)$~compute the KL distance between the type and the $H$ candidate PMFs (hypotheses) $p_h$, $h=1,\dots,H$; $(iii)$~decide for the hypothesis yielding the minimum KL distance, as prescribed by the right-hand 
side of~(\ref{eq:argmaxargmin}).

\begin{figure}
\centering 
\includegraphics[width =240pt]{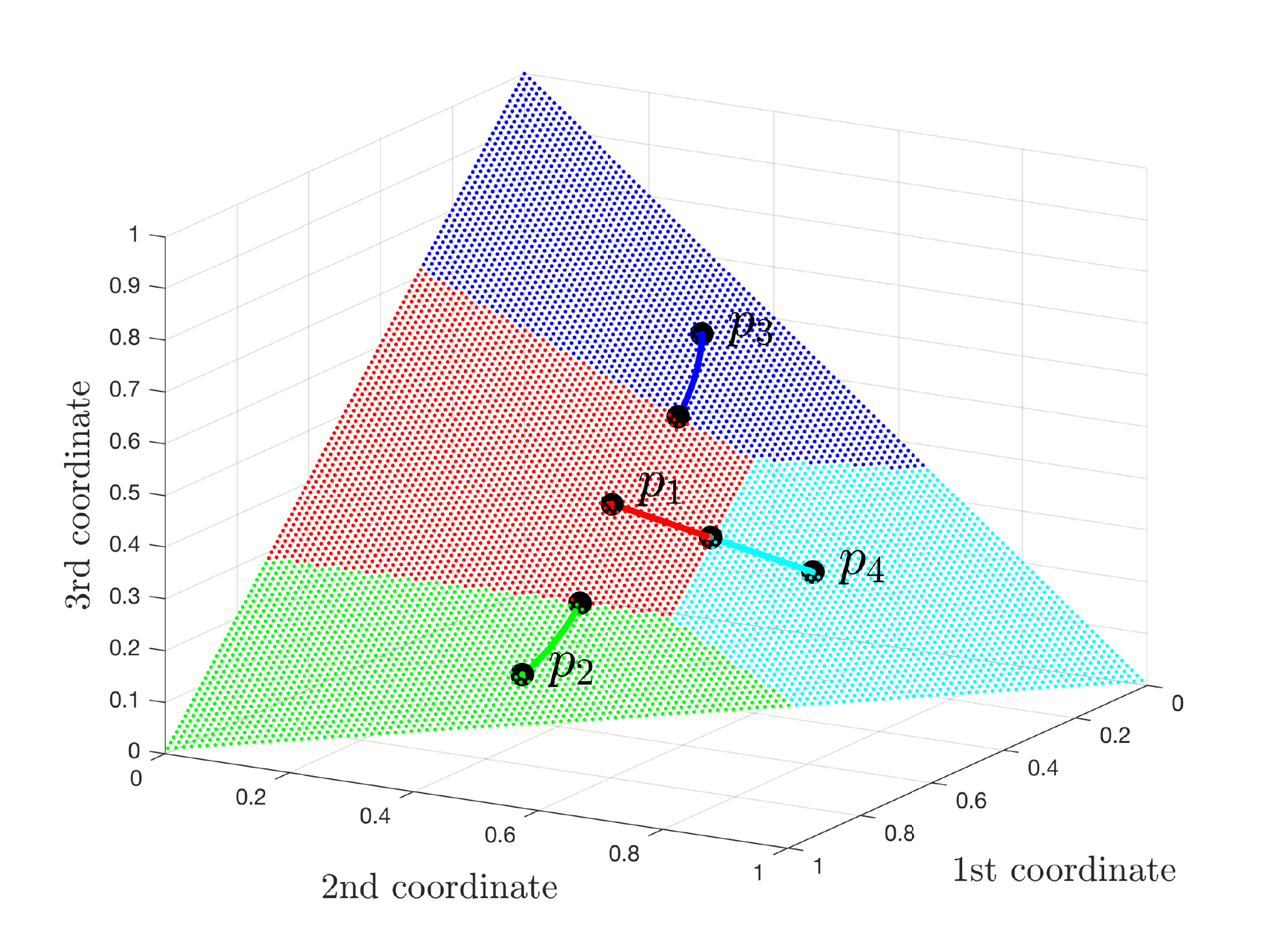}
 \caption{Geometry of multi-hypothesis decisions in an example where there are $H=4$ hypotheses, represented by the four PMFs $p_1,\dots,p_4$, with cardinality $|\cA|=3$. The probability simplex is a triangle of the three-dimensional space and each point in the simplex is a PMF, whose entries (``coordinates'') represent the probabilities of the three letters of the alphabet. The four hypotheses partition the simplex into cells (colored regions) having the PMFs relative to the hypothesis as centroids  (bullets). The colored lines show the geodesics connecting the centroids to their closest cell border.}
      \label{fig:simplexIA}
\end{figure}

The probability simplex is thus partitioned into $H$ convex decision cells with piecewise straight line borders, having as centroids the PMFs representing the $H$ hypotheses. 
This is illustrated\footnote{The example in Fig.~\ref{fig:simplexIA} refers to Eq.~(\ref{eq:PMFexample1}) of Sec.~\ref{sec:computerIA}, with $\alpha=0.25$.} in Fig.~\ref{fig:simplexIA}, for $H=4$ hypotheses and alphabet of size~3.
Also shown in the figure are the geodesics connecting the centroids to their nearest cell border using the KL ``metric''. 
The length of the $h$-th geodesic measures the asymptotic (large $n$) exponential rate at which the best error probability conditional to hypothesis $\cH_h$ approaches zero. 
The overall error probability $P(e)$ in~(\ref{eq:pe}) scales exponentially at rate given by the length of the shortest geodesic, which is 
the minimum Chernoff information~\cite{CT2} between the ${H}\choose{2}$ pairs of hypotheses, see~\cite{Westover} for details.

\subsection{Algorithm for Multi-Hypothesis Decision}
\label{sec:LMSmulti}
Of course, in the case of adaptive implementations, there is hardly any hope in obtaining 
an asymptotically optimal decision maker similar to~(\ref{eq:hath}), because 
decision performance must be sacrificed to ensure adaptation.  
However, the previous arguments and geometrical interpretation suggest a suitable decision procedure, as follows. First,
as soon as a new observation $\bx(i)$ is available at time~$i$, the agent computes $H$ log-likelihoods: 
\begin{align}
\bd^{(h)}(i)=\log p_h(\bx(i)), \quad h=1,\dots,H.
\label{eq:ll}
\end{align} 
Note that, since data $\{ \bx(i)\}_{i\ge 1}$ are conditionally IID, so are the $\{\bd^{(h)}(i)\}_{i\ge 1}$ with respect to the time index $i$, for each fixed~$h$. 
Then, employing $\{\bd^{(h)}(i)\}_{i\ge 1}$, the agent implements~$H$ parallel iterates of the LMS algorithm for decision shown in~(\ref{eq:LMSdet}), one for each of the $H$ log-likelihoods. 
We refer to these parallel iterates as the \emph{branches} of the algorithm. Let us denote by $\bw^{(h)}(n)$ the output of the $h$-th branch at time $n$. 
The decision of the agent at time $n$ is in favor of the hypothesis corresponding to the branch yielding the maximum output:
\begin{align}
&\widehat \bh(n) =\arg\max_h \bw^{(h)}(n).
\label{eq:decag}
\end{align}

The rationale is as follows. Let us assume that $\cH_{h^\star}$ is the true hypothesis. 
For $n\gg1$ we know from~(\ref{eq:EVAR}) that $\bw^{(h)}(n)$, as function of $n$, oscillates around its mean value $\E_{h^\star} \log p_h(\bx)$, 
and these oscillations can be made as small as desired by using a sufficiently small value of $\mu$. 
By the strong law of large numbers~\cite{billingsley-book2}, with probability one:
\begin{align}
\lim_{n\rightarrow \infty} \frac 1 n \sum_{i=1}^n \log p_h(\bx(i))  = \E_{h^\star} \log p_h(\bx),  
\end{align}
which shows that for large $n$ the optimal decision maker shown in~(\ref{eq:decag})  
approaches the mean value of $\bw^{(h)}(n)$.
Recall also that $D(p_{h^\star} || p_h) \ge 0$ with equality if, and only if, $p_{h^\star}(a)=p_h(a)$, $\forall a \in \cA$, see e.g.,~\cite{CT2}. This implies that 
under hypothesis $\cH_{h^\star}$, the mean value of the $h^\star$-th branch output of the LMS algorithm is given by the negative entropy of $p_{h^\star}$, and the mean values of all others $H-1$ branch outputs are smaller.

The picture obtained by Chebyshev inequality is equally informative~\cite{papoulis}.
For $\mu \ll 1$, the ${h^\star}$-th branch output of the LMS algorithm~(\ref{eq:LMSdet}) --- the one corresponding to the true hypothesis $\cH_{h^\star}$ --- lies with probability at least $(1-\epsilon) \in (0,1)$ in a neighborhood of size $\approx \sqrt{\mu \VAR_{h^\star} \bd^{({h^\star})}/(2\epsilon)} $ of the negative entropy $\E_{h^\star} \log p_{h^\star}(\bx)$. The $h$-th output, $h\neq {h^\star} $, lies with probability at least $(1-\epsilon)$ in a neighborhood of size $\approx \sqrt{\mu \VAR_{h^\star} \bd^{(h)}/(2\epsilon)}$ of $\E_{h^\star} \log p_h(\bx)<\E_{h^\star} \log p_{h^\star}(\bx)$.

In summary, the LMS algorithm in~(\ref{eq:LMSdet}) can be used for the decision problem at hand to learn the decision statistic. At the same time, should the hypothesis abruptly change, the algorithm would benefit from its inherent adaptation properties and the algorithm output will start approaching the decision statistics for the new hypothesis. Of course, this adaptation property comes at the cost of some sub-optimal decision performance in steady-state, i.e., when a hypothesis is in force for infinitely long time.
The fundamental tradeoff between learning and adaptation is controlled by the step size~$\mu$.

\section{Agent Network: Diffusion Rule for Decision}
\label{sec:ATC}

Having motivated the usage of the LMS algorithm~(\ref{eq:LMSdet}), fed by $\{\bd^{(h)}(i)\}_{i\ge1}$ given in~(\ref{eq:ll}), in the context of multi-hypothesis decision problems, we now proceed to extend
the algorithm to multi-agent scenarios, paralleling the well-established generalization of~(\ref{eq:LMS0}) to multi-agent networks  in the context of estimation 
problems~\cite{KhawatmiSP17}.

Consider hence a network of $S$ agents interconnected by a graph structure with nodes representing agents and undirected edges representing connections between them. 
An example of such network is shown in Fig.~\ref{fig:net}. 
One approach to regulate the interactions of each agent $k\in\{1,\dots,S\}$ with other agents in the network is represented by the so-called ATC 
(adapt-then-combine) diffusion rule~\cite[Eq.~(49b)]{SayedprocIEEE}:
for $h=1,\dots, H$, set $\bw^{(h)}_k(0)=0$ and, for $i\ge1$,
\begin{subequations} \begin{align}
&\bv_k^{(h)}(i)=\bw_k^{(h)}(i-1)+\mu \left [\bd_k^{(h)}(i)-\bw_k^{(h)}(i-1) \right ], \label{eq:old1}\\
\label{eq:old2} &\bw_k^{(h)}(i)= \sum_{\ell=1}^S a_{k \ell} \bv_{\ell}^{(h)}(i), 
\end{align}\label{eq:old}\end{subequations}
where 
\begin{align}
\bd_k^{(h)}(i)=\log p_h(\bx_k(i)), \quad h=1,\dots,H.
\label{eq:ll2}
\end{align}
The quantities $\bw_k^{(h)}(i)$, $h=1,\dots,H$, will be referred to as the \emph{status} of agent $k$ at time $i$.
In~(\ref{eq:old}), the LMS algorithm shown in~(\ref{eq:LMSdet}) is employed in~(\ref{eq:old1}) to compute an intermediate value $\bv_k^{(h)}(i)$, which is subsequently combined with the intermediate values from the neighbors of agent~$k$ through~(\ref{eq:old2}), to yield the updated status of the agent. This combination uses the scalars $\{a_{k\ell}\}_{\ell=1}^S$ that represent the weight by which information flowing from agent~$\ell$ to agent~$k$ is scaled.  These weights satisfy
\beq
a_{k\ell}\geq 0,\quad \sum_{\ell\in \cI_k} a_{k\ell}=1,
\label{eq:akl}
\eeq
where $\cI_k$ denotes the set of neighbors connected to agent~$k$ by an edge in the graph. Note that~$\cI_k$ includes~$k$ itself: $k \in \cI_k$.
We set $a_{k\ell}=0$ if $\ell \not \in \cI_k$, implying that only local interactions are allowed. 
For simplicity of notation, $a_{kk}$ is denoted by $a_k$, which is assumed strictly positive.
Organizing the weights in matrix form results in a right-stochastic matrix with non-negative entries $A=[a_{k\ell}]$.

Other algorithms have been proposed in the literature of multi-agent adaptive systems.
One notable example is the CTA (combine-then-adapt) diffusion rule obtained by switching the order of the steps in~(\ref{eq:old})~\cite[Eq.~(49a)]{SayedprocIEEE}:
for $h=1,\dots,H$,
\begin{subequations} \begin{align}
&\bv_k^{(h)}(i-1)= \sum_{\ell=1}^S a_{k \ell} \bw_{\ell}^{(h)}(i-1), \label{eq:cons1} \\
&\bw_k^{(h)}(i)=\bv_k^{(h)}(i-1)+\mu [\bd_k^{(h)}(i)-\bv_k^{(h)}(i-1)], \label{eq:cons2}
\end{align}\label{eq:cons}\end{subequations}
but there is some advantage to using the ATC shown in~(\ref{eq:old}) rather than the CTA or other similar variants, as discussed in~\cite{TowficChenSayedIT2016}.
If in~(\ref{eq:cons2}) a diminishing step-size is employed in place of a constant value, i.e., $\mu=\mu(i)\to0$ for $i\to \infty$, we obtain an instance of \emph{running consensus} scheme studied in~\cite{asymptotic-rc,running-cons,Bracaetal-Pageconsensus}, see e.g.,~\cite[Eq.~(1)]{running-cons}.
On the other hand, if we replace the second occurrence of $\bv_k^{(h)}(i-1)$ at the right-hand side of~(\ref{eq:cons2}) by $\bw_k^{(h)}(i-1)$, we obtain
an algorithm belonging to the class of \emph{consensus} adaptive schemes, see e.g.,~\cite[Eq.~(46c)]{SayedprocIEEE}. 
In general, diffusion algorithms have 
wider stability ranges and improved performance over consensus schemes, both for constant and diminishing step-sizes,
in view of the existing asymmetry in the adaptation step of consensus implementations,
as explained in~\cite{SayedprocIEEE,SayedNOW2014,TowficChenSayedIT2016}.
Moreover, diminishing step-sizes limit the adaptation ability of the network, because the weights assigned to new measurements tend to vanish, 
which limits the use of diminishing step-sizes in dynamic environments~\cite{MaranoSayedIT19,BracaetalIT}.  

For these reasons, in the following we focus on the ATC diffusion scheme with constant step size shown in~(\ref{eq:old}).
A detailed study of the decision performance of multi-agent networks employing a scheme similar to~(\ref{eq:old}) has been carried out 
in~\cite{BracaetalIT,MattaSIPN16},
and a variation thereof, designed for networks with limited link capacity, can be found in~\cite{MaranoSayedIT19}. 
In all these references, the simplest case of only two hypotheses has been considered, and the arguments in Secs.~\ref{sec:multi} and~\ref{sec:LMSmulti}
suggest how to generalize the results to a multi-hypothesis scenario. 
We do not go into the details of these generalizations.
Instead, we now focus on the case of \emph{multi-task adaptive networks}, for which 
there is recent literature addressing estimation problems~\cite{ChenJSTSP17,KhawatmiSP17,NassifSP17,NassifSPL19,KhawatmiSP2019}
but only limited if hardly any studies are available for decision problems similar to those considered in the next two sections. 


\section{Decisions with Informed Agents}
\label{sec:IA}

\subsection{Problem Formulation}

We now consider the case of multi-task networks for decision problems, in which agents of the network belong to different \emph{clusters}.
The observations made by agents belonging to the same cluster are conditionally IID, similarly to the case of single-task networks addressed so far. However, observations made by agents belonging to different clusters are conditionally independent but may not be identically distributed. Note that the conditioning is with respect to the \emph{global} state of nature, i.e., given the observation models of all clusters.

\begin{figure}
\centering 
\includegraphics[width =190pt]{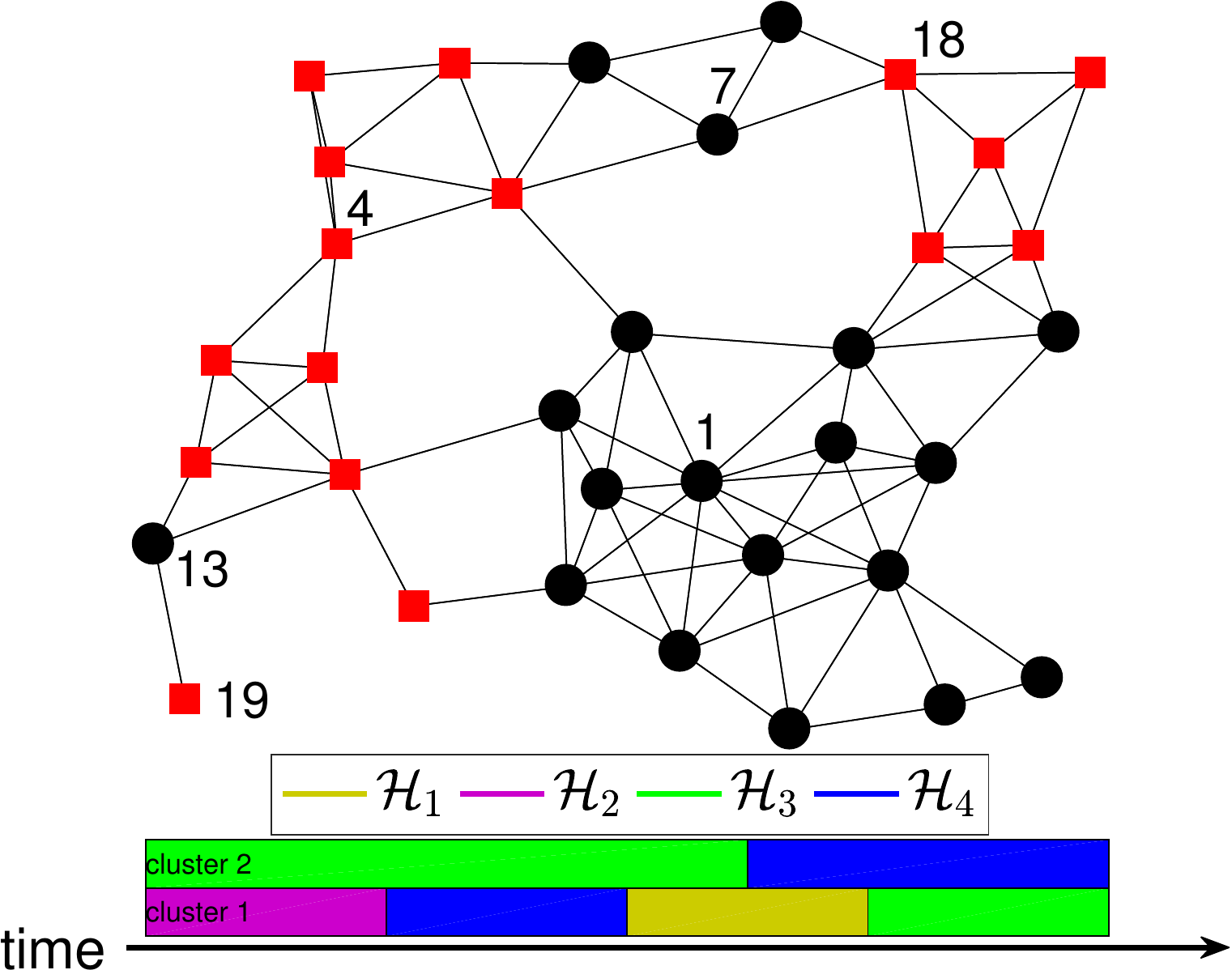}
 \caption{\emph{Top:} An example of multi-task network. Agents belonging to different clusters are shown by different symbols/colors, here black circles and red squares. 
In multi-task networks agents have neighbors, but only those in the same cluster are \emph{effective} neighbors. For instance, agent 1 has 11 neighbors and all are effective, while agent 13 has 4 neighbors and only one (itself) is effective.
 \emph{Bottom:} At a given time agents of different clusters work under different states of nature (hypotheses $\cH_1, \cH_2,\dots$), which change in time asynchronously and unpredictably, as shown by the bands of different colors.}
      \label{fig:net}
\end{figure}

An example will help clarify the addressed scenario. Consider the network in Fig.~\ref{fig:net} made of $S=35$ agents grouped in two clusters, and suppose that there are $H=4$ 
possible states of nature $\cH_1,\dots,\cH_4$, characterized by the four PMFs $p_1,\dots,p_4$. 
In Fig.~\ref{fig:net} different symbols/colors represent agents belonging to different clusters.
Suppose the observations made by agents in the first cluster, say those denoted by squares, are initially drawn from a certain PMF, say $p_2$, and at some later time the distribution 
changes to $p_4$, then to $p_1$ and then to $p_3$. Likewise, suppose the observations made by agents in the second cluster (circles) are initially drawn from a PMF $p_3$,
which at some later time changes to $p_4$. The succession of the states of nature for the two clusters are illustrated by the bands in the bottom part of Fig.~\ref{fig:net} using different colors for different hypotheses, as shown in the legend. The task of the network is to make decisions about the state of nature at each instant of time, for each agent.

\emph{Agents are informed}, meaning that the ensemble of all PMFs 
$p_1, \dots,p_H$, corresponding to the possible state of nature $\cH_1, \dots, \cH_H$ are known to all agents. However, 
\emph{agents do not know the cluster that they belong to}, neither do they know when and how the state of nature changes for each cluster.
In fact, we address the challenging scenario in which the states of nature change in time in an unpredictable and uncontrollable way, also asynchronously with respect to the various clusters, as illustrated in the bottom part of Fig.~\ref{fig:net}. It is not excluded that, at any given time, the observations made by different clusters are drawn from the same probability distribution.

Due to the complete lack of knowledge about the succession of states of nature, agents have no hope to learn from the past some rule of succession: 
the network is faced with a multi-hypothesis test among $H$ equally likely alternatives, at each time instant, for each agent.

\begin{figure}
\centering 
\includegraphics[width =100pt]{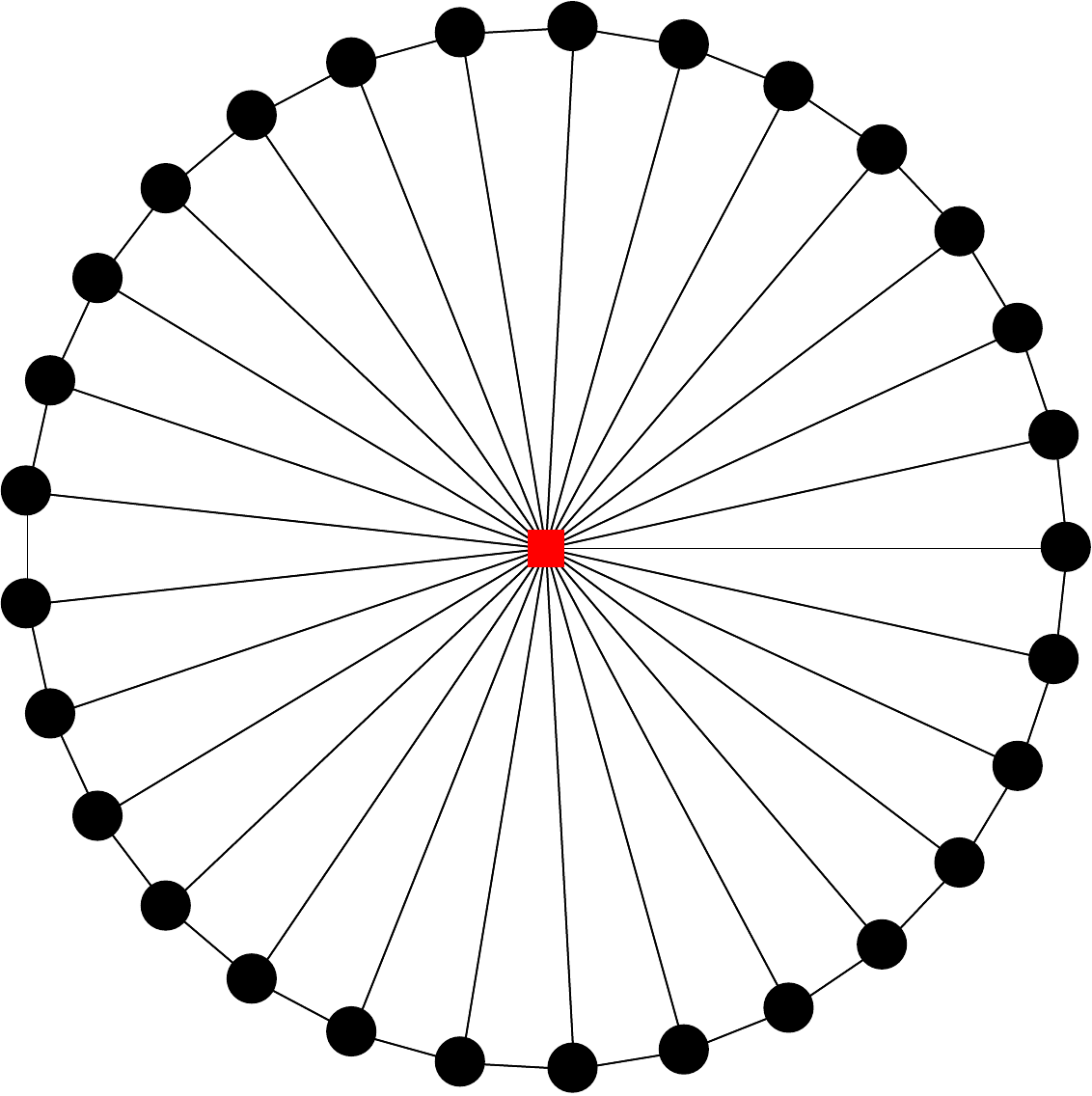}
 \caption{A circular network in which the central agent belongs to a cluster and all other agents to a different cluster. This ad-hoc configuration is used to illustrate the challenges arising in multi-task networks, where the presence of non-effective neighbor agents is potentially catastrophic.}
      \label{fig:netcircle}
\end{figure}

\subsection{Challenges of Multi-Task Scenario}

In single-task networks, we have seen that one suitable decision criterion for multi-hypothesis decision problems amounts to selecting the branch yielding the maximum output among the $H$ parallel branches of the ATC procedure shown in~(\ref{eq:old}), fed by the values $\{\bd_k^{(h)}(i)\}_{i\ge1}$ defined in~(\ref{eq:ll2}).
A natural generalization to the multi-task scenario would be as follows. First, let us define the \emph{effective} neighbors $\cE_k$ of agent $k$ as those agents in $\cI_k$ 
whose cluster is the same as that of agent~$k$. Second, define at each time $i$ the set $\widehat{\bcE}_k(i)$ of \emph{estimated} effective neighbors of agent $k$, as those agents from $\cI_k$ whose local decision at time $i-1$ is the same as the decision of agent~$k$: 
\begin{align}
&&\widehat{\bcE}_k(i) \dfz \{\ell \in  \cI_k \, : \, &\arg\max_h \bw_\ell^{(h)}(i-1) \nonumber \\
&&&  = \arg\max_h \bw_k^{(h)}(i-1) \}.
\label{eq:hatNno}
\end{align}
The set $\widehat{\bcE}_k(i)$ is random and contains the indexes of the agents in $\cI_k$ that at time $i$ are believed to belong to the same cluster as agent $k$.  
Computation of $\widehat{\bcE}_k(i)$ requires a modest communication burden among agents, to deliver their decisions to the neighbors. 

Then, the combination step~(\ref{eq:old2}) of the ATC procedure for agent $k$ is modified to include only the neighbor agents belonging to $\widehat{\bcE}_k(i)$:
the right-stochastic combination matrix~$A$ with nonnegative entries becomes random and time-varying:
\beq
\bA(i)=[\ba_{k\ell}(i)], \qquad \textnormal{with } \ba_{k\ell}(i)= 0   \textnormal{ for }  \ell \not \in \widehat{\bcE}_k(i).
\label{eq:comb(i)}
\eeq
The entries $\{a_k\}_{k=1}^S$ on the main diagonal of $\bA(i)$ are nonrandom, do not vary in time, are strictly positive, and represent design parameters.

In the absence of any reason to distinguish between the agents belonging to $\widehat \bcE_k(i) \setminus \{k\}$, a meaningful choice for the combination matrix is
\begin{align}
\ba_{k\ell}(i)= \begin{cases}
0,  & |\widehat \bcE_k(i)|>1 \; \;\& \; \; \ell \not \in \widehat \bcE_k(i), \\ 
\frac{1-a_k}{|\widehat \bcE_k(i)|-1}, & |\widehat \bcE_k(i)|>1   \; \;\& \; \; \ell \in \widehat \bcE_k(i) \setminus \{k\},  \\
a_k, &  |\widehat \bcE_k(i)|>1   \; \; \& \; \;   \ell=k,   \\ \\
0, & |\widehat \bcE_k(i)|=1  \; \; \& \; \;  \ell \not = k,    \\  
1, & |\widehat \bcE_k(i)|=1 \; \; \& \; \;  \ell=k,
\end{cases}
\label{eq:A(i)typical}
\end{align}
where the last two lines of~(\ref{eq:A(i)typical}) apply when no agent in $\cI_k$ makes the same decision as agent $k$,
in which case the combination step of the ATC rule is void, and $\bw_k^{(h)}(i)= \bv_{\ell}^{(h)}(i)$.

\begin{figure}
\centering 
\includegraphics[width =210pt]{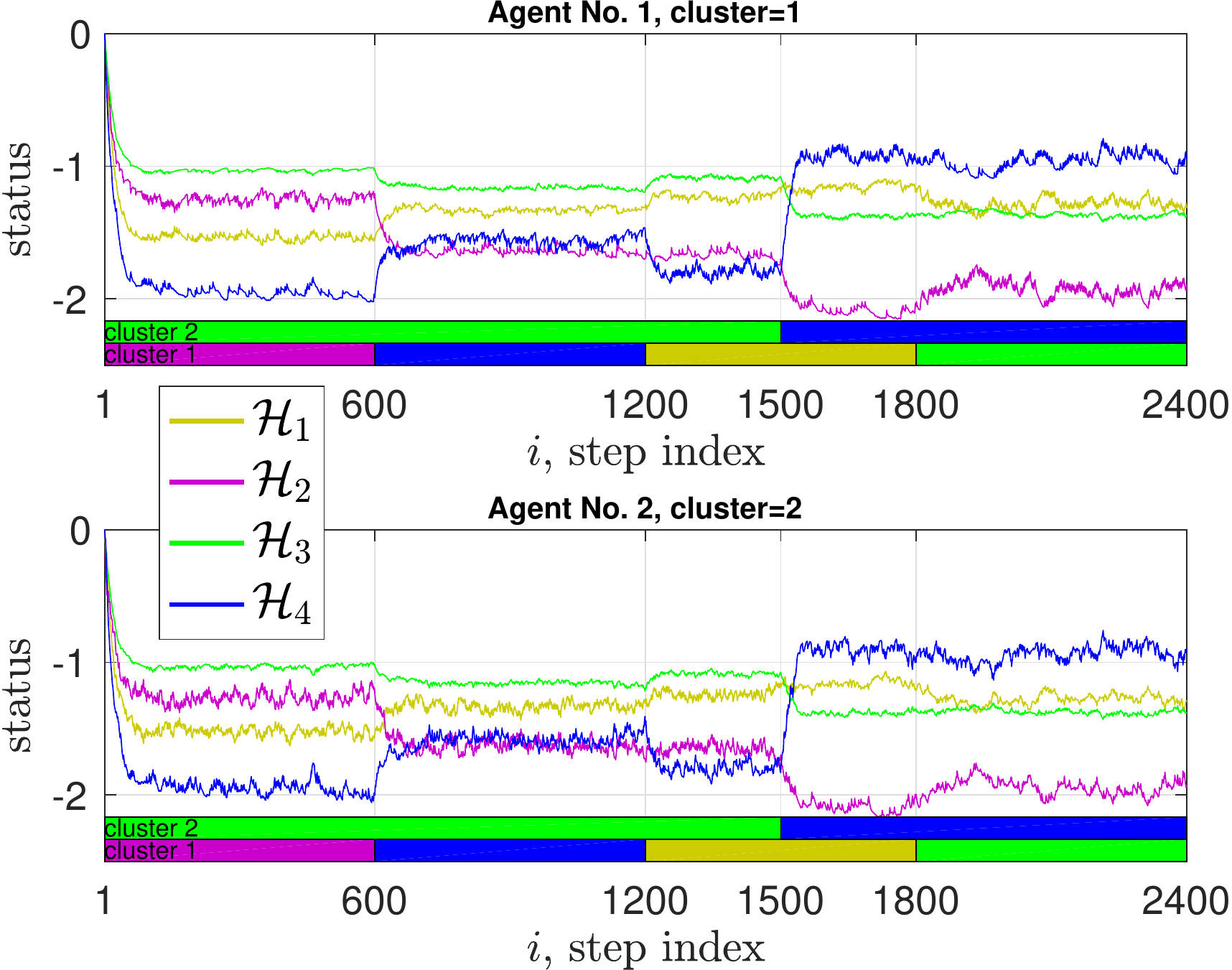}
 \caption{A single realization of the status of the agent in the center (top) and of an agent on the circumference (bottom) of the circular network shown in Fig.~\ref{fig:netcircle}. The status is composed of four elements and the largest of these elements determines the decision of the agent. Note in the top panel that agent~1 behaves as if it belonged to  cluster~2, which is due to the presence of many non-effective neighbors. Without a careful design of the diffusion mechanism, the presence of neighbors can impair the agent's decision ability.}
      \label{fig:open1}
\end{figure}

Although~(\ref{eq:hatNno}) and~(\ref{eq:comb(i)}) appear as natural choices, we must be careful.
In many applications, most neighbor agents belong to the same cluster, because the cluster membership depends on the geographic position of the agents. 
However, we want to address the general case in which the cluster membership is not structured and it may happen, for instance, that all neighbors of a given agent
belong to a cluster different from that of the agent. This situation is particularly challenging, and a system design based on~(\ref{eq:hatNno}) and~(\ref{eq:comb(i)}) may not be sufficient. 

To illustrate the problem, consider the network shown in Fig.~\ref{fig:netcircle} with $S=30$ agents, in which the central agent (square) belongs to cluster 1 and all other agents on the circumference (circles)  belong to cluster 2. 
Let $\mu=0.05$, and let the matrix $\bA(i)$ be that shown in~(\ref{eq:A(i)typical}), with $a_k=0.5$ for all agents.
Suppose $H=4$ and assume that the four PMFs corresponding to $\cH_1,\dots,\cH_4$, are
\begin{align}
&p_1=[0.6 \; 0.3 \; 0.1], && p_2=[0.1 \; 0.1 \; 0.8], \\ 
& p_3=[0.4 \; 0.2 \; 0.4], && p_4=[0.1 \; 0.8 \; 1].
\label{eq:PMFs}
\end{align}
Implementing algorithm~(\ref{eq:old}), with $\{\bd_k^{(h)}(i)\}_{i\ge1}$ given in~(\ref{eq:ll2}), we obtain the agent status $\bw_k^{(h)}(i)$, with $h=1,\dots,4$. For $1\le i \le 2400$, one realization of this status is shown by the colored curves in the top panel of Fig.~\ref{fig:open1} for $k=1$ (central agent), and in the bottom panel for $k=2$ (one of the agents on the circumference). 
The colors of the curves refer to the four elements $\bw_k^{(h)}(i)$, $h=1,\dots,4$ of the status, the largest of which determines the decision of agent $k$ at time~$i$, see~(\ref{eq:decag}).

The global state of nature is shown in the bottom part of the two panels of Fig.~\ref{fig:open1}. It is seen that agent 1, belonging to cluster~1, tends to behave as if it belongs to cluster 2. This can be explained by considering that agent 1 is surrounded by many agents of cluster 2 and therefore in the presence of occasional wrong decisions, agent~1 combines its intermediate status with those of the surrounding agents. At that point, the status of agent~1 begins to be strongly influenced by that of its neighbors, with the disastrous consequences shown in Fig.~\ref{fig:open1}. 
In summary, in multi-task networks, with arbitrary cluster memberships and unpredictable changes in the state of nature, the presence of the neighbors, instead of providing a beneficial diversity, may prove to be catastrophic if not handled properly.

\begin{figure}
\centering 
\includegraphics[width =210pt]{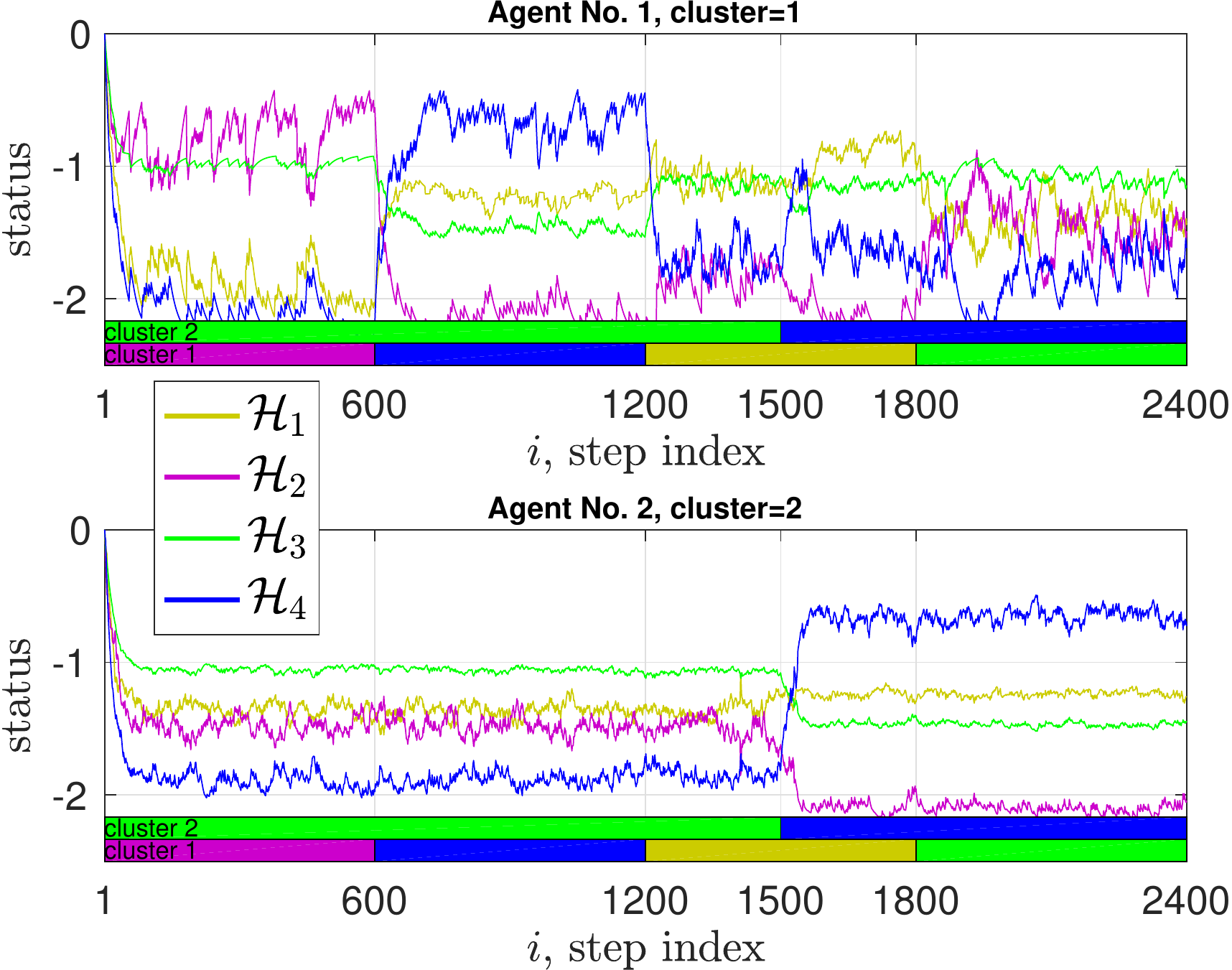}
 \caption{A single realization of the status of the agent in the center (top) and of an agent on the circumference (bottom) of the circular network shown in Fig.~\ref{fig:netcircle}, for the diffusion rule proposed in this paper for multi-task networks. Comparing this figure with Fig.~\ref{fig:open1}, we see that the decisions of agent 1 are now correct most of time.}
      \label{fig:open2}
\end{figure}

\subsection{Algorithm Design}

The aforementioned difficulties can be avoided by separating the status update from the mechanism used to estimate the cluster membership, at the cost of some additional local computation. Our approach is motivated by the formulation developed in~\cite{ZhaoSayed-SP2015} for estimation problems.

Let $\bz_k^{(h)}(i)$, $h=1,\dots,H$, be the status of agent $k$ at time~$i$ when the agent is assumed to be isolated from the network, namely, let us define $\bz_k^{(h)}(0)=0$, and for $i\ge 1$,
\begin{align}
\bz_k^{(h)}(i)=\bz_k^{(h)}(i-1)+\mu [\bd_k^{(h)}(i)-\bz_k^{(h)}(i-1)], \label{eq:mon}
\end{align}
where $\bd_k^{(h)}(i)$ is defined in~(\ref{eq:ll2}). Status~(\ref{eq:mon}) is used to update step-by-step the combination matrix $\bA(i)$, by defining the set
$\widehat \bcE_k(i)$ of estimated effective neighbors of agent $k$ as 
\begin{align}
&&\widehat \bcE_k(i) \dfz \{\ell \in  \cI_k \, : \, &\arg\max_h \bz_\ell^{(h)}(i-1) \nonumber \\
&&&  = \arg\max_h \bz_k^{(h)}(i-1) \},
\label{eq:hatN}
\end{align}
where~(\ref{eq:hatN}) employs $\bz_\ell^{(h)}(i)$
instead of $\bw_\ell^{(h)}(i)$, which was employed in~(\ref{eq:hatNno}). 
Clearly, now each agent updates two iterates: $\bz_k^{(h)}(i)$ and $\bw_k^{(h)}(i)$,
both driven by the same sequence $\{\bd_k^{(h)}(i)\}_{i\ge1}$ defined in~(\ref{eq:ll2}). The former evolves as described in~(\ref{eq:mon}) and characterizes  
agent~$k$ in isolation, while the latter evolves according to~(\ref{eq:old}), with $\bA(i)$
as in~(\ref{eq:comb(i)}), and $\widehat \bcE_k(i)$ given by~(\ref{eq:hatN}).

Returning to the example addressed in Fig.~\ref{fig:open1}, using the proposed modification~(\ref{eq:hatN}) in place of~(\ref{eq:hatNno}), we obtain
more comfortable results, as seen in Fig.~\ref{fig:open2} where the local decisions of agent 1 are now correct most of the time. 
The following algorithm summarizes the proposed ATC diffusion scheme for decision in multi-task environments, with reference to the generic $k$-th agent.

\begin{algorithm}[h]
\small{
\NoCaptionOfAlgo
\DontPrintSemicolon
\KwIn{ $\bd_k^{(h)}(i)=\log p_h(\bx_k(i))$, $h=1,\dots,H$, $i\ge 1$, see~(\ref{eq:ll2})} 
\KwOut{$\bw_k^{(h)}(i)$, $i\ge 1$, $h=1\dots,H$ \\ \hspace*{32pt} decision at time $i$: $\arg\max_{h} \bw_k^{(h)}(i)$} $\,$ \\
\textbf{Initialize:} $\bz_k^{(h)}(0)=0$, and $\bw_k^{(h)}(0)=0$, $h=1,\dots,H$ \\
\textbf{for $i=1,2,\dots,$ do} \\
\quad compute $\widehat \bcE_k(i)=\{\ell \in  \cI_k \, : \, \arg\max_h \bz_\ell^{(h)}(i-1) = $ \\
\hspace*{80pt} $\arg\max_h \bz_k^{(h)}(i-1) \}$, see~(\ref{eq:hatN}) \\
\quad using $\widehat \bcE_k(i)$, define $\bA(i)$ right-stochastic satisfying \\
\hspace*{50pt} $\ba_{k\ell}(i)= 0, \, \ell \not \in \widehat \bcE_k(i)$. E.g., use definition~(\ref{eq:A(i)typical}) \\
\quad \textbf{for $h=1,\dots,H$, do} \\
\qquad $\bz_k^{(h)}(i)=\bz_k^{(h)}(i-1)+\mu [\bd_k^{(h)}(i)-\bz_k^{(h)}(i-1)]$ \\
\qquad $\bv_k^{(h)}(i)=\bw_k^{(h)}(i-1)+\mu [\bd_k^{(h)}(i)-\bw_k^{(h)}(i-1)]$ \\
\qquad $\bw_k^{(h)}(i)= \sum_{\ell=1}^S \ba_{k \ell}(i) \bv_{\ell}^{(h)}(i)$ \\
\quad \textbf{end for} \\
\textbf{end for}
\\
\caption{\textbf{Algorithm IA: Decisions by Informed Agents}}
}
\end{algorithm}

\section{Event Detection with Partially-Informed Agents}
\label{sec:PIA}

\subsection{Problem Formulation}

A second decision problem that is of great relevance for practical applications is now introduced.
Suppose that the network is engaged in a binary detection task, as follows. Either observations made at all agents are IID and come from one and the same ``null'' model $\cH_0$, statistically characterized by PMF $p_0$, or some ``alternative'' event $\cH_1$ takes place, in which case data collected by agents are 
independent and drawn from \emph{unknown} PMFs, different from~$p_0$. We have two possible states of nature and we want to decide which is true between $\cH_0$ and $\cH_1$. We refer to this decision problem as that of partially-informed agents because we assume that agents of the network know the PMF $p_0$
but the statistical distributions of their measurements under the alternative hypothesis $\cH_1$ are unknown. 

As before, agents are grouped into clusters and, under~$\cH_1$,
agents of the same cluster collect IID data, while agents belonging to different clusters collect independent but possibly non-identically distributed data. 
We do not pose restrictions on the distributions active under $\cH_1$, except that they must be strictly positive over the same alphabet as $p_0$, and different
from $p_0$. In addition, the succession of states $\cH_0 \mapsto \cH_1 \mapsto \cH_0, \dots$ is arbitrary --- changing times are unknown and not probabilistically modeled --- and agents of each cluster may experience different alternative distributions under different ``$\cH_1$'' epochs. Changes in the state of nature are synchronous across the whole network. 

We do not assign a-priori probabilities to~$\cH_0$ and~$\cH_1$. 
The decision performance is measured in terms of type~I error probability (also called false alarm), which is the probability of making the wrong decision under $\cH_0$, and type~II error probability (miss detection), which is the probability of making the wrong decision under $\cH_1$. The reason is that when the presence of some ``alternative'' situation $\cH_1$ must be detected, which intervenes to modify the ``normal'' status of nature $\cH_0$, the two error events may have quite different meaning and impact.
In these situations, a sensible optimization criterion is the Neyman-Pearson formulation~\cite{Lehmann-testing3}: minimize type~II error probability subject to a constraint on type~I error probability.

\subsection{Algorithm Design}
\label{sec:AD2}

\subsubsection{Test Formulation}

In the informed agent decision problem, the rationale behind the design of the diffusion algorithm was founded on constructing a surrogate of the optimal log-likelihood decision statistic. 
In the case of partially informed agents, it is self-evident that the approach must be modified, because agents cannot compute anymore the log-likelihoods as in~(\ref{eq:ll2}). Then, our goal is to make the agent status an approximation of the \emph{type} (empirical PMF) of the observations, 
in light of the asymptotic optimality of the Hoeffding procedure~\cite{hoeffding}:
\begin{align}
D(t_{\bx_k(1:n)} || p_0) \test \gamma
\label{eq:Hoeffding}
\end{align} 
in testing a known $p_0$ ``against everything else''.
In~(\ref{eq:Hoeffding})  $t_{\bx_k(1:n)}$ denotes the type of the sequence $\bx_k({1:n})=[\bx_k(1), \bx_k(2),$ $\dots,\bx_k(n)]$ of observations at agent $k$, $D(\cdot || \cdot)$ denotes KL divergence~\cite{CT2}, $\gamma$ is a suitable threshold level, and the qualification asymptotic refers to $n\to\infty$ in $\bx_k(1:n)$, assuming that such sequence is made of IID components. 
%
%

Using~(\ref{eq:Hoeffding}) as guideline, the design of the diffusion algorithm for multi-task networks is now addressed. Without loss in generality, let $\cA=\{1,\dots,M\}$ be the observation alphabet, so that $\bx_k(i) =m$, for some $m=1,\dots,M$. 
In place of~(\ref{eq:ll2}), we define at any agent $k$, for $i \ge 1$,
\begin{align}
\bd_k^{(m)}(i)= \left \{
\begin{array}{ll}
1, \quad \bx_k(i)=m, \\
0, \quad \bx_k(i)\neq m,
\end{array}
\right .
\label{eq:d_PI}
\end{align}
namely, $\bd_k^{(m)}(i)$ is the indicator of the event $\{ \bx_k(i)=m\}$.

Let us consider the following ATC-like rule: 
set $\bw_k^{(m)}(0)=1/M$, for $m=1\dots,M$ (uniform initial guess). Then, 
for $m=1,\dots, M$, $i\ge1$, $k=1\dots,S$:
\begin{subequations} \begin{align}
&\bv_k^{(m)}(i)=\bw_k^{(m)}(i-1)+\mu [\bd_k^{(m)}(i)-\bw_k^{(m)}(i-1)], \label{eq:ATC_PI_1}\\
&\bw_k^{(m)}(i)= \sum_{\ell=1}^S a_{k \ell} \bv_{\ell}^{(m)}(i), \label{eq:ATC_PI_2}
\end{align}\label{eq:reold}\end{subequations}
where $\bd_k^{(m)}(i)$ is now defined by~(\ref{eq:d_PI}). 
The decision made by agent $k$ at time $i$ is 
\begin{align}
D(\bw_k(i) || p_0) \test \gamma,
\label{eq:Hoeffding2}
\end{align} 
where $\bw_k(i)=[\bw_k^{(1)}(i),\dots,\bw_k^{(M)}(i)]^T$ and where $\gamma$ is chosen to satisfy the type~I error constraint.

\subsubsection{Challenges}
\label{sec:dacit}
Suppose for a moment that the coefficients $a_{k\ell}$ in~(\ref{eq:ATC_PI_2}) are constant in time and satisfy~(\ref{eq:akl}), which means that all neighbors of agent $k$
are included in the combination step~(\ref{eq:ATC_PI_2}), regardless of the cluster that they belong to.
In such case, when $\cH_0$ is in force for a time long enough and $\mu \ll1$, using relationships~(\ref{eq:EVAR}) we expect that $\bw_k^{(m)}(i)$ approximates
the occurrence probability of symbol $m$ under probability model~$p_0$. 
Likewise, under hypothesis $\cH_1$, $\bw_k^{(m)}(i)$ approximates a weighted combination of the probabilities of symbol $m$ under the different observation models
of all agents in the network. 
This would be problematic because such weighted combination could be close to the probability of symbol $m$ under $p_0$, and this ``balancing effect'' impairs the decision ability of the network. This argument shows that also in the case of partially informed agents, any agent $k$ needs to implement
a mechanism aimed at selecting its \emph{effective} neighbors, thus avoiding potentially dangerous influences from neighbor agents belonging to different clusters.

\subsubsection{Addressing the Challenges} In the same spirit of the approach {proposed in~\cite{ZhaoSayed-SP2015} and pursued in Sec.~\ref{sec:IA}, we introduce the iterate $\bz_k(i)$ that uses only the data collected by agent $k$. This status is used to obtain a measure of distance between the status of agent $k$ and the status of its neighbors. Only when this distance is small enough, the neighbor agent is included in the combination step of~(\ref{eq:reold}). In formulas, similarly to~(\ref{eq:mon}):
 $\bz_k^{(m)}(0)=1/M$, $m=1,\dots,M$, and for $i\ge 1$,
\begin{align}
\bz_k^{(m)}(i)=\bz_k^{(m)}(i-1)+\mu [\bd_k^{(m)}(i)-\bz_k^{(m)}(i-1)], \label{eq:mon2}
\end{align}
where $\bd_k^{(m)}(i)$ is given by~(\ref{eq:d_PI}). 

Exploiting~(\ref{eq:mon2}) we define a time-varying combination matrix $\bA(i)$, by introducing the set
$\widehat \bcE_k(i)$ of estimated effective neighbors of agent $k$ as 
\begin{align}
\widehat \bcE_k(i) \dfz \{\ell \in  \cI_k \, : \, &\| \bz_k(i-1)-\bz_\ell(i-1) \| < \delta \},
\label{eq:hatN2}
\end{align}
where in computing the Euclidean norm, $\bz_k(i-1)$ is regarded as an $M$-vector with entries
$\bz_k^{(1)}(i-1)$, $\bz_k^{(2)}(i-1)$, \dots, $\bz_k^{(M)}(i-1)$. Similarly for $\bz_\ell(i-1)$.
In~(\ref{eq:hatN2}) the threshold level~$\delta$ is a design parameter. 
Combination matrix $\bA(i)$ is defined as a right-stochastic matrix of nonnegative entries with the property $\ba_{k\ell}=0$ for $\ell \not \in 
\widehat \bcE_k(i)$, where $\widehat \bcE_k(i)$ is given by~(\ref{eq:hatN2}).
One typical choice for $\bA(i)$ is shown in~(\ref{eq:A(i)typical}).

At the cost of some repetition, Algorithm~PIA summarizes the decision/diffusion procedure
with reference to the generic $k$-th agent.

\begin{algorithm}[h]
\small{
\NoCaptionOfAlgo
\DontPrintSemicolon
\KwIn{ $\bd_k^{(m)}(i)$, $m=1,\dots,M$, $i\ge 1$, see~(\ref{eq:d_PI})  \\  \hspace*{28pt}  two thresholds: $\gamma$ (see output below), and $\delta$, see~(\ref{eq:hatN2})}
\KwOut{$\bw_k^{(m)}(i)$, $i\ge 1$, $m=1\dots,M$ \\ \hspace*{32pt} decision at time $i$: $D \big(\bw_k(i) || p_0\big) {\tiny{ \test}} \gamma $} $\,$ \\
\textbf{Initialize:} $\bz_k^{(m)}(0)=\frac 1 M$, and $\bw_k^{(m)}(0)=\frac 1 M$, $m=1,\dots,M$ \\
\textbf{for $i=1,2,\dots,$ do} \\
\quad set $\widehat \bcE_k(i)=\{\ell \in  \cI_k  : \| \bz_k(i-1)-\bz_\ell(i-1) \| < \delta \}$ [(\ref{eq:hatN2})] \\
\quad using $\widehat \bcE_k(i)$, define $\bA(i)$ right-stochastic satisfying \\
\hspace*{50pt} $\ba_{k\ell}(i)= 0, \, \ell \not \in \widehat \bcE_k(i)$. E.g., use definition~(\ref{eq:A(i)typical}) \\
\quad \textbf{for $m=1,\dots,M$, do} \\
\qquad $\bz_k^{(m)}(i)=\bz_k^{(m)}(i-1)+\mu [\bd_k^{(m)}(i)-\bz_k^{(m)}(i-1)]$ \\
\qquad $\bv_k^{(m)}(i)=\bw_k^{(m)}(i-1)+\mu [\bd_k^{(m)}(i)-\bw_k^{(m)}(i-1)]$ \\
\qquad $\bw_k^{(m)}(i)= \sum_{\ell=1}^S \ba_{k \ell}(i) \bv_{\ell}^{(m)}(i)$ \\
\quad \textbf{end for} \\
\textbf{end for}
\\
\caption{\textbf{Algorithm PIA: Event Detection by Part.\ Infor.\ Agents}}
}
\end{algorithm}

\section{Approximate Statistical Characterization}
\label{sec:approx}

In this section, the statistical characterization of the agents' status for the two decision problems of informed and partially informed agents is derived under the simplifying assumption that the diffusion algorithm employs the exact combination matrix, $A$. 
Namely, we let $\widehat{\bcE}_k(i) =\cE_k$ in~(\ref{eq:comb(i)}), as if the clustering operation would be made without errors.
To provide a unified treatment of the two decision problems, we use the index $r=1\dots,R$, to denote, respectively: 
\begin{align}
\hspace*{-5pt}\begin{cases} h=1\dots,H, & \textnormal{for the case of informed agents,} \\ m=1,\dots,M, & \textnormal{for case partially informed agents.}\end{cases}
\end{align}
By straightforward algebra the following explicit expression for $\bw_k^{(r)}(i)$ can be derived:
\begin{subequations} \label{eq:ta1}\begin{align} 
& \bw_k^{(r)}(i)=\sqrt{\mu}\sum_{j=1}^i \bt_k^{(r)}(i,j),  \label{eq:ta11}  \\
&\hspace*{-8pt} \bt_k^{(r)}(i,j) \dfz \hspace*{-3pt} \sqrt{\mu} (1-\mu)^{j-1} \sum_{\ell=1}^S b_{k\ell}(j) \, \bd_\ell^{(r)}(i-j+1), \label{eq:ta12}
\end{align}\end{subequations}%
where $ b_{k\ell}(j)$ is the $(k,\ell)$-entry of matrix $B(j)=A^j$.
By stacking the quantities in~(\ref{eq:ta12}) for $r=1,\dots,R$,
into a vector: 
\begin{align} \label{eq:vect}
\bq_k(i,j)\dfz \left [ \bt_k^{(1)}(i,j), \dots, \bt_k^{(R)}(i,j) \right ]^T,
\end{align}
we obtain a triangular array of vectors as follows:
{\small \begin{align} \label{eq:array}
\begin{matrix}
\bq_k(1,1) \\
\bq_k(2,1) & \bq_k(2,2) \\
\bq_k(3,1) & \bq_k(3,2) & \bq_k(3,3) \\
\dots & \dots & \dots &\dots \\
\bq_k(i,1) & \bq_k(i,2) & \bq_k(i,3) & \dots & \bq_k(i,i) \\
\dots & \dots & \dots &\dots & \dots & \dots 
\end{matrix}
\end{align}}%
Henceforth, $\bd_k^{(r)}(i)$, whose definition for the two decision problems is given in~(\ref{eq:ll2}) and~(\ref{eq:d_PI}), is abbreviated as $\bd_k^{(r)}$ whenever the value of the index $i$ is immaterial.

\begin{theorem} \label{th}
In the limit $i \to \infty$ followed by $\mu\to 0$, the sum $\sum_{j=1}^i \big [ \bq_k(i,j)- \E\bq_k(i,j) \big ]$ of the zero-mean version of the $i$-th row of
the array~(\ref{eq:array}) converges in distribution to a multivariate zero-mean Gaussian vector with $R \times R$ covariance matrix~$\beta_k(A) \, \Lambda_k $,
where the entries of matrix $\Lambda_k$ are given by
\begin{align}
&[\Lambda_k]_{r_1r_2}= \E \Big [ \big (\bd_k^{(r_1)} -\E \bd_k^{(r_1)} \big )  \big (\bd_k^{(r_2)} -\E \bd_k^{(r_2)} \big )\Big ], \label{eq:lambda}
\end{align}
and the scalar $\beta_k(A)$ is defined as
\begin{align}
\beta_k(A)=\lim_{\mu\to 0} \sum_{j=1}^\infty \sum_{\ell=1}^S \mu (1-\mu)^{2j-2} \, b_{k\ell}^2(j), \label{eq:beta}
\end{align}
and satisfies
\begin{align}
\frac{1}{2S} \le \beta_k(A) \le \frac 1 2. \label{eq:bounds}
\end{align}
\end{theorem}

\begin{IEEEproof}
See Appendix~\ref{app:proof1}.
\end{IEEEproof}

\begin{corollary} \label{cor}
Let 
\begin{align}\bz_k^{(r)}(i)=\sqrt{\mu}\sum_{j=1}^i \bt_k^{(r)}(i,j), \label{eq:zincor}
\end{align} 
where $\bt_k^{(r)}(i,j)$ is defined as in~(\ref{eq:ta12}) with $b_{k\ell}(j)$ replaced by one if $k=\ell$ and zero otherwise. Then,
in the limit $i \to \infty$ followed by $\mu\to 0$, the sum $\sum_{j=1}^i \big [ \bq_k(i,j)- \E\bq_k(i,j) \big ]$ from~(\ref{eq:array})
converges in distribution to a multivariate zero-mean Gaussian vector with $R \times R$ covariance matrix~$\Lambda_k/2$,
where $\Lambda_k$ is given in~(\ref{eq:lambda}).
\end{corollary}

\begin{IEEEproof}
The result trivially follows from
Theorem~\ref{th}, because using the zero/one value of $b_{k\ell}(j)$ indicated in the Corollary, from~(\ref{eq:beta}) one immediately gets $\beta_k(A) = 1/2$.
\end{IEEEproof}

Denoting by $\bw_k^{(h)}$ and $\bz_k^{(h)}$ the steady-state values of the algorithm outputs obtained by letting $i\to \infty$ in~(\ref{eq:ta1}) and~(\ref{eq:zincor}),
respectively, the previous results allow us to make the following approximation\footnote{$\cN_R(c,\Sigma)$ denotes a jointly Gaussian distribution of size $R$, with mean~$c$ and covariance matrix~$\Sigma$. When $R=1$ this notation is simplified to $\cN(c,\Sigma)$.} for $\mu \ll 1$:
\begin{align} 
& \left [\bw_k^{(1)}, \dots, \bw_k^{(R)}\right ]^T \sim \cN_R \Big ( c_k , \mu \, \beta_k(A) \, \Lambda_k \Big ), \label{eq:on1}\\
& \left [\bz_k^{(1)}, \dots, \bz_k^{(R)}\right ]^T \sim \cN_R \Big( c_k ,  \frac { \mu \, \Lambda_k}{2} \Big), \label{eq:on2}
\end{align}
where 
$c_k=  \big[\E \bd_k^{(1)},  \dots,\E \bd_k^{(R)} \big]^T$.
In particular, for $r\in\{1,\dots,R\}$,
\begin{align} 
&\bw_k^{(r)} \sim \cN \Big(  \E\bd_k^{(r)} , \; \mu\, \beta_k(A) \; \V \bd_k^{(r)} \Big), \label{eq:wandzw}\\
&\bz_k^{(r)} \sim \cN \Big(  \E\bd_k^{(r)},  \; \mu\, \frac 1 2  \; \V\bd_k^{(r)} \Big). \label{eq:wandzz}
\end{align}

Note that, if~$p_{h^\star}$ is the actual distribution for agent $k$, the expectation in~(\ref{eq:lambda}) is computed under $p_{h^\star}$, 
and the explicit expression for the expectations appearing in~(\ref{eq:on1})-(\ref{eq:wandzz}) for informed and partially informed agents are, respectively,
\begin{align} 
\begin{cases}
\E\bd_k^{(h)} = \sum_{a\in \cA} p_{h^\star}(a) \log p_h(a), & h=1,\dots,H,\\
\E\bd_k^{(m)} = p_{h^\star}(m), & m=1,\dots,M.
\end{cases}
\end{align}

From~(\ref{eq:on1})-(\ref{eq:wandzz}) we see that the variance-reduction factor $\beta_k(A)$ quantifies the beneficial effect of collaboration among agents, and accounts for the topology of the network. If $A$ is the identity matrix, which models isolated agents, $\beta_k(A)=1/2$ for all~$k$. To the other extreme, 
if all entries of $A$ are equal to $1/S$, then $\beta_k(A)=1/(2S)$ for all~$k$, which corresponds to a fully connected single-task network with all agents belonging to the same cluster.
These special values of matrix $A$ yield the bounds shown in~(\ref{eq:bounds}).

\section{Computer Experiments}
\label{sec:comp}

\subsection{Informed Agents}
\label{sec:computerIA}

The diffusion algorithm for informed agents provides inherent adaptation properties to the network. 
This is confirmed, for instance, by Fig.~\ref{fig:open2}, where we see that agents promptly react to changes in the state of nature. 
A natural question at this point is to inquire about the decision performance of the network. 
We now investigate by computer experiments the decision properties of the adaptive system at steady-state, namely when the state of nature is constant for a long period of time.

With reference to the network shown in Fig.~\ref{fig:net} (but ignoring now the colored bands in the bottom panel), let us consider the following setup: $\mu=0.05$, 
combination matrix $\bA(i)$ as in~(\ref{eq:A(i)typical}) with $a_k=0.5$ for all agents, $\widehat \bcE_k(i)$ given by~(\ref{eq:hatN}),  and 
$H=4$ hypotheses, characterized respectively by the PMFs:
\begin{align}
\begin{array}{l}
p_1=[p_{11}, \, p_{12}, \; p_{13}],  \\  p_2=[p_{11}+\alpha, \; p_{12}, \; p_{13}-\alpha], \\ 
p_3=[p_{11}-\alpha, \; p_{12}, \; p_{13}+\alpha], \\ p_4=\Big [p_{11}-\frac\alpha 2, \; p_{12}+\alpha, \; p_{13}-\frac \alpha 2	 \Big],
\end{array}
\label{eq:PMFexample1}
\end{align}
wherein $p_{11}=p_{12}=p_{13}=1/3$. Note that the larger is $\alpha\ge 0$, the more ``different'' are the PMFs, while for $\alpha=0$ the four hypotheses
collapse into one and the same distribution.

The results of computer experiments using $n=1000$ steps (namely, each agent $k$ takes the decision after that the state of nature has been constant for $1000$ steps of the algorithm) and 5000 Monte Carlo runs are shown in the top plot of Fig.~\ref{fig:simul01} for the case in which agents of cluster~1 are under hypothesis~$\cH_4$ (PMF $p_4$)
and agents of cluster~2 are under hypothesis $\cH_1$ (PMF $p_1$). The bottom bottom plot of  Fig.~\ref{fig:simul01} shows the case in which agents of cluster~1 are under hypothesis $\cH_3$ and agents of cluster~2 are under hypothesis $\cH_2$.

Figure~\ref{fig:simul01} shows the error probability conditioned to the global state of nature (i.e., given the observation models of all clusters) 
$P(e\,|\,\textnormal{s.o.n.})$, at the four agents 1, 13, 4, and 18, where the decisions are made according to~(\ref{eq:decag}).
As seen in Fig.~\ref{fig:net},  agent~1 belongs to cluster~1 and has 11 neighbors (recall that the number of neighbors includes the agent itself) and 11 effective neighbors, namely
all its neighbors belong to its own cluster. We see that the error probability at agent~1 decreases rapidly with $\alpha$, and this can be explained by the substantial diversity gain
obtained by exchanging its status with that of the neighbors. Agent~13, which is also of cluster~1, has 4 neighbors and only 1 of these agents (itself) belongs to cluster 1. We expect worse performance with respect to agent~1, because there is no diversity gain is this case. Indeed, while for very small values of $\alpha$ the error probability is essentially the same of that experienced by agent~1, such probability decreases now more slowly when $\alpha$ grows.

Consider next agent~4, belonging to cluster~2, which has~5 neighbors all of which are from cluster~2, and let us compare its error probability with that of
agent~18, belonging to cluster~2, which has~5 neighbors but only~3 are effective. As expected, agent 4 performs better than agent~18.
Note that for very small values of $\alpha$, the error probability essentially depends only on the cluster of the agent, while at large $\alpha$ the error probability seems smaller for agents with larger number of effective neighbors. This is further indication that the network exploits properly the presence of neighbors of the same cluster.
Comments to the bottom plot of Fig.~\ref{fig:simul01} are similar. 

\begin{figure}
\centering 
\includegraphics[width =210pt]{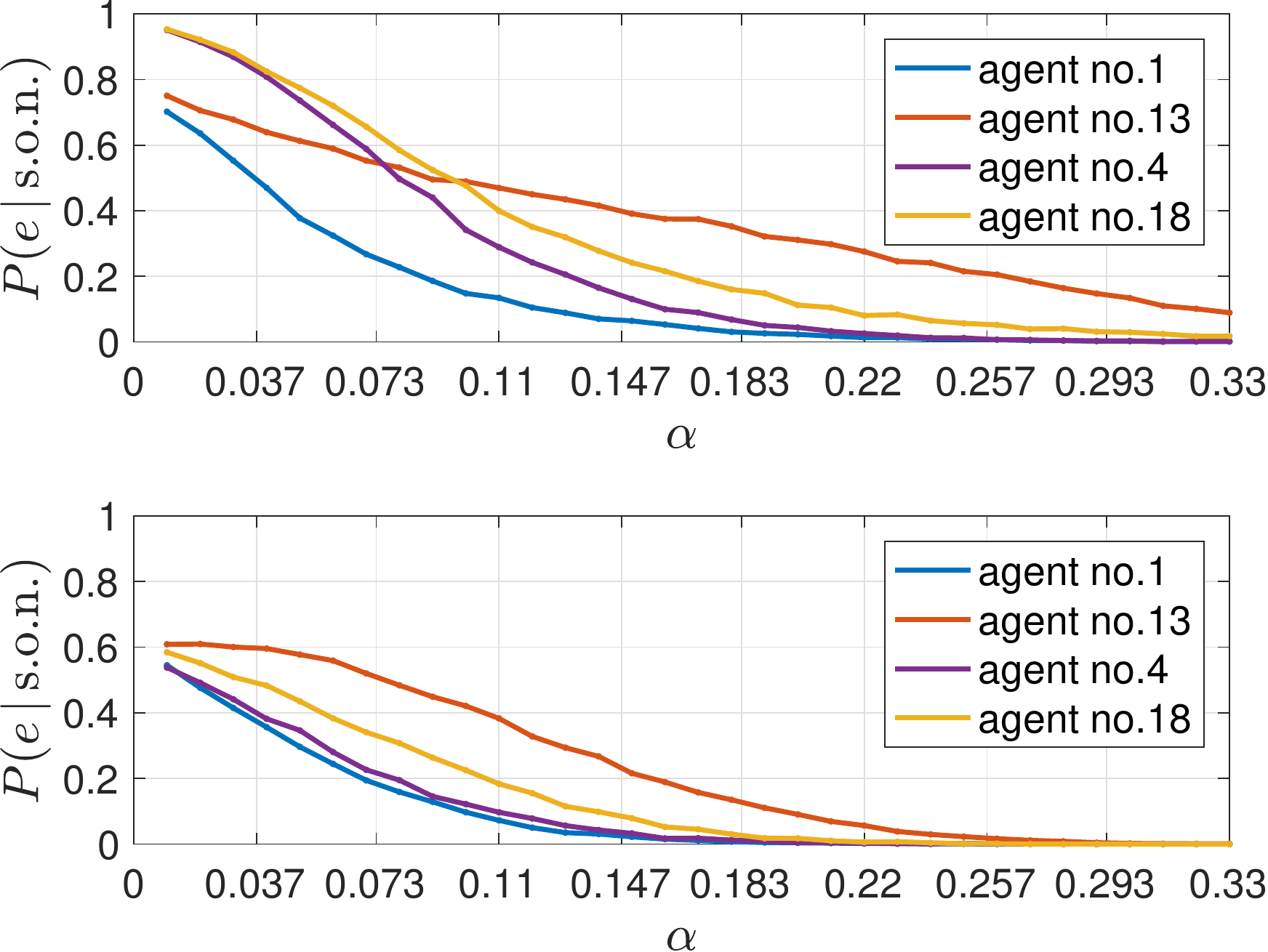}
 \caption{Steady-state error probability given the state of nature $P(e \, | \,\textnormal{s.o.n.})$ in function of $\alpha$ for four agents of the network shown in the top part of Fig.~\ref{fig:net}. The observation model is shown in~(\ref{eq:PMFexample1}). Recall that agents~1 and~13 belong to cluster~1, while agents~4 and~18 to cluster 2. In this example $\mu=0.05$, $\bA(i)$ is shown in~(\ref{eq:A(i)typical}) with $a_k=0.5$, for all $k$, and $\widehat \bcE_k(i)$ is given by~(\ref{eq:hatN}). \emph{Top:} Agents belonging to cluster~1 work under hypothesis~$\cH_4$
and agents of cluster~2 under hypothesis~$\cH_1$. \emph{Bottom:} Agents belonging to cluster~1 work under hypothesis~$\cH_3$
and agents of cluster~2 under hypothesis~$\cH_2$.}
      \label{fig:simul01}
\end{figure}

Next, we exploit the results of Sec.~\ref{sec:approx} to provide approximate upper and lower bounds for the error probability of agent~$k$.
Recall that $\bw_k^{(h)}$ and $\bz_k^{(h)}$, without the time index enclosed in parenthesis, denote quantities at steady-state.
Even knowing the joint distribution of $\{\bw_k^{(h)}\}_{h=1}^H$, computing an analytical expression for the error probability of agent~$k$ requires that we compute
$\P\{\arg\max_h \bw_k^{(h)} \not = h^\star\}$, where $p_{h^\star}$ is the data distribution relative to agent $k$. This computation is not straightforward.
It is more convenient to exploit the theoretical results of Sec.~\ref{sec:approx} to design a simplified setup for computer simulations, 
which achieves remarkable saving in terms of execution times with respect to the plain simulation of the diffusion algorithm IA.
Specifically, we implement standard Monte Carlo runs to estimate $\P\{\arg\max_h \bz_k^{(h)} \not = h^\star\}$, by exploiting 
the joint statistical characterization of $\{\bz_k^{(h)}\}_{h=1}^H$ shown in~(\ref{eq:on2}). This estimate is taken as an approximate upper bound of the actual performance.
Likewise, exploiting the joint characterization of $\{\bw_k^{(h)}\}_{h=1}^H$ given in~(\ref{eq:on1}), $\P\{\arg\max_h \bw_k^{(h)} \not = h^\star\}$ is estimated by Monte Carlo counting and is taken as 
an approximate lower bound of the actual performance, because of the assumption of knowing the connection matrix~$A$.

In Fig.~\ref{fig:simulboundsIA} the same example of Fig.~\ref{fig:simul01} is considered. The error probability for agents 1 and 4 is shown
along with the approximate lower and upper bounds. 
Note that, by adjusting in an ad-hoc manner the variance-reduction factor~$\beta_k(A)$ in~(\ref{eq:beta}), within the range shown in~(\ref{eq:bounds}),
the lower bound becomes a close approximation, as shown by the small circles in the figure. 
This observation hints at future approaches for obtaining reliable approximations, beyond the derived bounds.

Other agents of the network in Fig.~\ref{fig:open1} show similar behavior, with the exception of agents with a large number of non-effective neighbors relative to the effective ones, 
such as agents~7, 13, 19. For these agents, the error probability may be larger than the error probability of the agents themselves operating in isolation,
and the upper bound can be violated. In these circumstances, our performance prediction is of limited utility and a different approach should be pursued.

\begin{figure}
\centering 
\includegraphics[width =210pt]{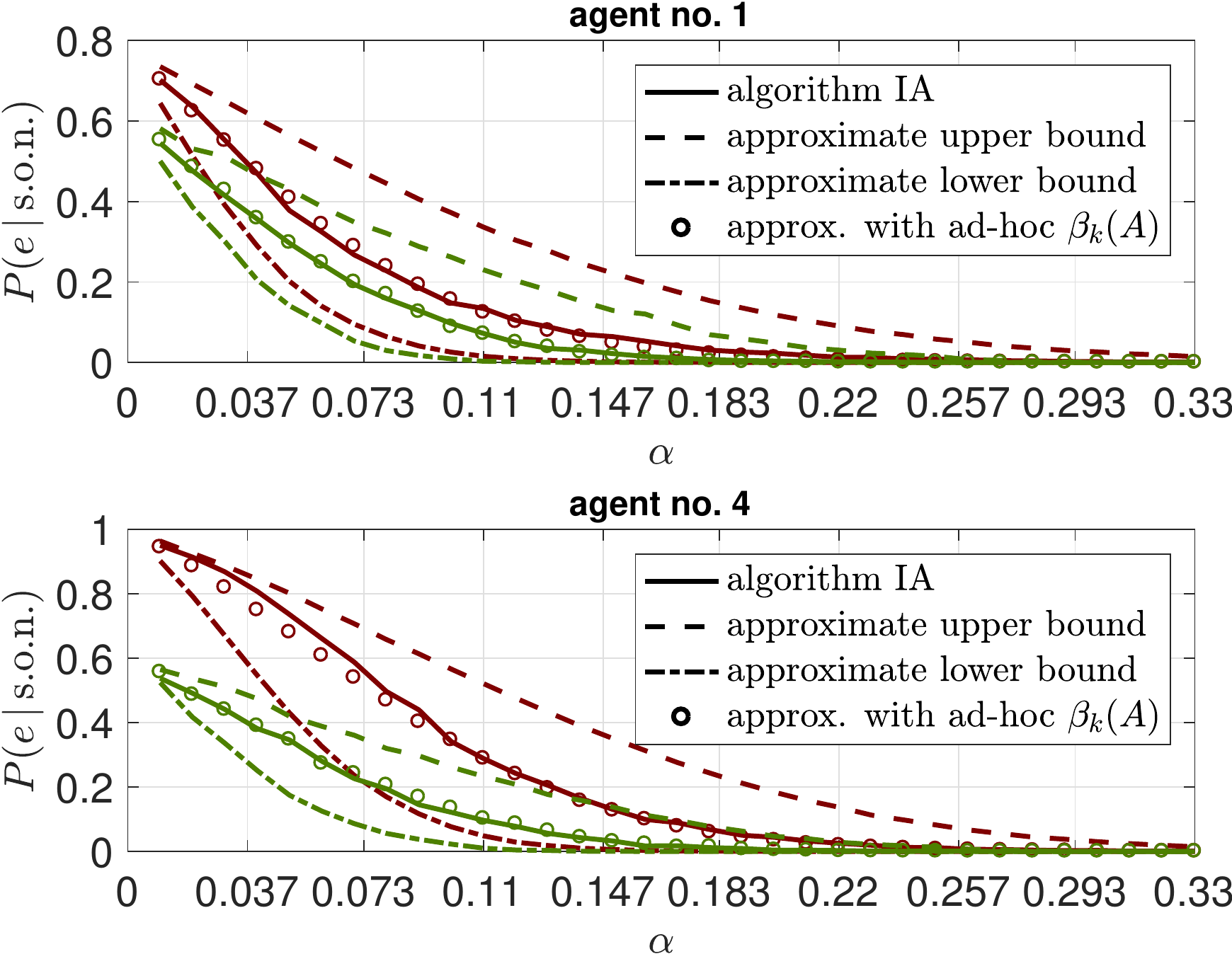}
 \caption{The same example of Fig.~\ref{fig:simul01} is considered. \emph{Red (top panel in Fig.~\ref{fig:simul01}):} Agents belonging to cluster~1 work under hypothesis~$\cH_4$
and agents of cluster~2 under hypothesis~$\cH_1$. \emph{Green (bottom panel in Fig.~\ref{fig:simul01}):} Agents belonging to cluster~1 work under hypothesis~$\cH_3$
and agents of cluster~2 under hypothesis~$\cH_2$. Solid curves refer to the plain simulation of Algorithm~IA. Dashed and dash-and-dotted curves show
the upper and lower bound, respectively, obtained by exploiting the theoretical results of Sec.~\ref{sec:approx}. The small circles are obtained from the lower bound
by adjusting ad-hoc the value of $\beta_k(A)$.}
      \label{fig:simulboundsIA}
\end{figure}

\subsection{Partially Informed Agents}

Consider next the decision properties of the adaptive system for partially-informed agents, at steady-state.
Let us refer again to the network shown in Fig.~\ref{fig:net} (ignoring the colored bands in the bottom panel). The simulation setup is as follows: $\mu=0.05$, 
the combination matrix $\bA(i)$ is that shown in~(\ref{eq:A(i)typical}) wherein $a_k=0.5$ for all agents, and $\widehat \bcE_k(i)$ is given by~(\ref{eq:hatN2}). 
Hypothesis $\cH_0$ is modeled by PMF
\begin{align}
&p_0=[p_{01}, \, p_{02}, \; p_{03}], 
\label{eq:PMFexample2H0}
\end{align}
with $p_{01}=p_{02}=p_{03}=1/3$. Under $\cH_1$, we assume the following distributions
\begin{align}
& p_1=[p_{01}+\alpha, \; p_{02}, \; p_{03}-\alpha], \label{eq:PMFexample2_1} \\ 
& p_2=[p_{01}-\alpha, \; p_{02}, \; p_{03}+\alpha],  \label{eq:PMFexample2_2} \\ 
& p_3=\Big [p_{01}-\frac\alpha 2, \; p_{02}+\alpha, \; p_{03}-\frac \alpha 2	 \Big].
\label{eq:PMFexample2_3}
\end{align}
Note that the distributions in~(\ref{eq:PMFexample2H0})-(\ref{eq:PMFexample2_3}) are exactly those shown in~(\ref{eq:PMFexample1}), but with different subscripts.
Recall also that the distributions under $\cH_1$ are unknown to the agents.
By means of $5000$ Monte Carlo runs, we estimate the threshold $\gamma$ in~(\ref{eq:Hoeffding2}) such that type~I error probability equals $10^{-1}$, and then estimate the correspondent type~II error probability by $500$ Monte Carlo runs. In both cases, the agents' decisions are taken after $n=1000$ time steps during which the the state of nature remains constant.

\begin{figure}
\centering 
\includegraphics[width =210pt]{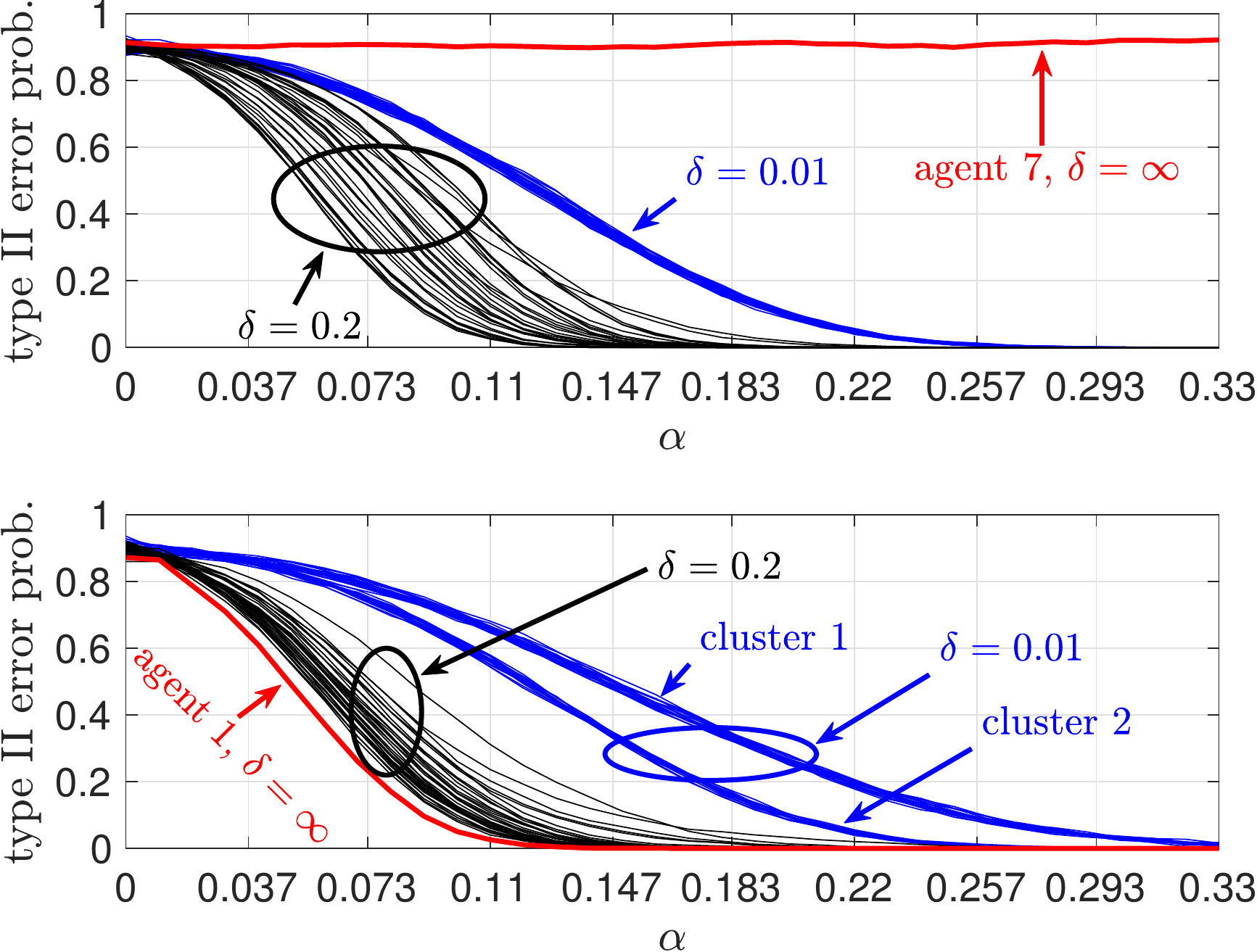}
 \caption{Steady-state type~II error probability in function of $\alpha$, for the 35 agents of the network shown in the top part of Fig.~\ref{fig:net}.
 The observation model is shown in~(\ref{eq:PMFexample2H0})-(\ref{eq:PMFexample2_3}). In this figure type~I error probability is set to $10^{-1}$, 
the step-size is $\mu=0.05$, and the combination matrix $\bA(i)$ is that shown in~(\ref{eq:A(i)typical}) with $a_k=0.5$ for all $k$, and $\widehat \bcE_k(i)$ given by~(\ref{eq:hatN2}).
Curves have been smoothed for better rendering.
\emph{Top:} Observations for agents of cluster 1 are drawn from $p_2$, and those for cluster~2 from $p_1$. 
The 35 curves in black refer to $\delta=0.2$. The 35 curves in blue almost exactly superimposed refer to $\delta=0.01$. Also shown in red is the performance of agent 7 with $\delta=\infty$.
\emph{Bottom:} Observations made by agents of cluster~1 are drawn from~$p_3$, and those made by agents of cluster~2 from~$p_2$. Curves in black refer to $\delta=0.2$, curves in blue refer to $\delta=0.01$. Also shown in red is the performance of agent 1 with $\delta=\infty$.}
      \label{fig:simulPIA01}
\end{figure}

The performance of the agents is shown in Fig.~\ref{fig:simulPIA01}. Consider first the top panel, in which we assume that agents of cluster~1 collect data drawn from PMF~$p_2$,
and agents of cluster~2 from~$p_1$. The curves in black refer to all 35 agents, with the threshold appearing in~(\ref{eq:hatN2}) given by $\delta=0.2$, a value chosen 
empirically after some trials and errors. 
We see that, for $\alpha\to0$, type~II error probability equals one minus type~I error probability, as it must be. As~$\alpha$ grows, type~II error probability decreases and approaches zero for large values of~$\alpha$; this monotonic behavior is expected because the larger is~$\alpha$ the more ``different'' are the distributions under $\cH_0$ 
and~$\cH_1$. 

The same dependence on $\alpha$ is observed for $\delta=0.01$, see curves in blue. 
As expected, for so small values of $\delta$, agents behave as if they were in isolation, that is, $\bv_k^{(m)}(i) \approx \bw_k^{(m)}(i) \approx \bz_k^{(m)}(i)$, see~(\ref{eq:reold}) and~(\ref{eq:mon2}). In this case, from the curves in blue, we see that the all agents perform similarly, which 
can be explained by noting that $D(p_0||p_1)=D(p_0||p_2)$ or, in other words, agents of the two clusters are faced with ``similarly difficult'' decision problems.
On the other hand, for large values of~$\delta$ the agents tend to interact with all their neighbors, both effective and not effective, and their status 
approximates an average of that of the neighboring agents.  Under $\cH_1$, 
due to the symmetry between the PMFs in~(\ref{eq:PMFexample2_1}) and~(\ref{eq:PMFexample2_2}), it may happen that the agent status approaches the PMF $p_0$ 
given in~(\ref{eq:PMFexample2H0}).
An example of this phenomenon is shown by the curve in red in the top panel of Fig.~\ref{fig:simulPIA01}, which is the performance of agent 7 when $\delta$ is so large that the inequality in~(\ref{eq:hatN2}) is always verified, implying $\widehat \bcE_k(i)=\cI_k$ for all $i$. This is a manifestation of the the balancing effect mentioned in Sec.~\ref{sec:dacit}.
We see that the choice of $\delta$ is critical for ensuring good decision performance at steady-state.

Consider next the bottom panel of Fig.~\ref{fig:simulPIA01}, where it is assumed that observations made by agents of cluster~1 are drawn from~$p_3$ and observations made by agents of cluster~2 from~$p_2$. The 35 curves in black show the type-II error probability of the agents for $\delta=0.2$. The behavior is qualitatively similar to that in the top panel. 
The 35 curves in blue refer to $\delta=0.01$ and represent the performance of isolated agents. As expected, when operating in isolation, agents of cluster~1 perform similarly (curves are superimposed), as do agents of cluster~2. The latter (with PMF $p_2$) perform slightly better than the former (PMF $p_3)$, which can be intuitively explained by noting that $D(p_0||p_2)>D(p_0||p_3)$.

\begin{figure}
\centering 
\includegraphics[width =210pt]{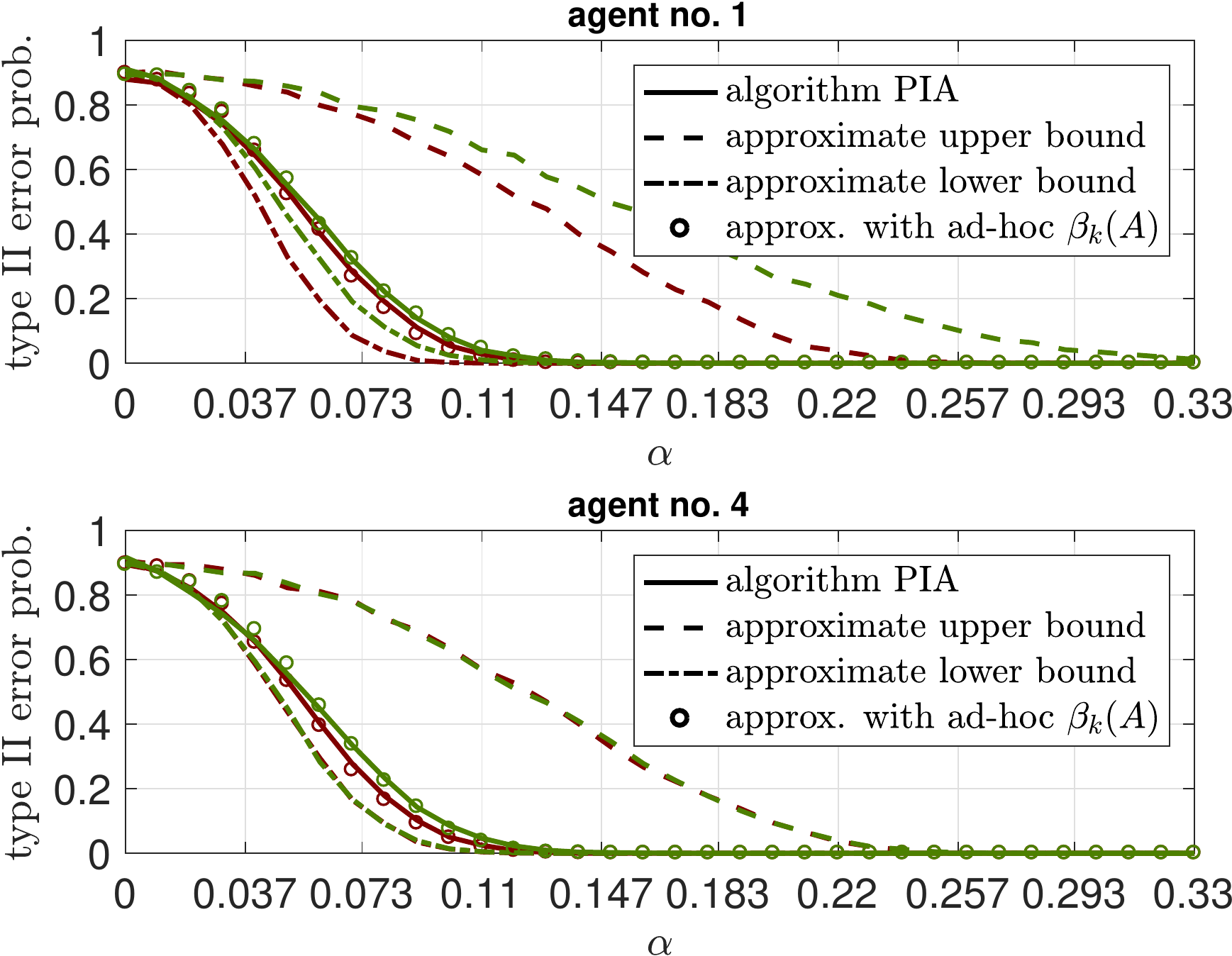}
 \caption{Same example as in Fig.~\ref{fig:simulboundsIA}, using $\delta=0.2$. 
 \emph{Red (top panel in Fig.~\ref{fig:simulPIA01}):} Observations for agents of cluster 1 are drawn from $p_2$, and those for cluster~2 from $p_1$. 
\emph{Green (bottom panel in Fig.~\ref{fig:simulPIA01}):} Observations made by agents of cluster~1 are drawn from~$p_3$, and those made by agents of cluster~2 from~$p_2$.
Solid curves (smoothed for better rendering) refer to the plain simulation of Algorithm~PIA. Dashed and dash-and-dotted curves show
the upper and lower bound, respectively, obtained by exploiting the theoretical results of Sec.~\ref{sec:approx}. The small circles are obtained from the lower bound
by adjusting ad-hoc the value of $\beta_k(A)$.}
      \label{fig:simulboundsPIA}
\end{figure}

Due to the structure of the PMFs there is no danger that a weighted combination of $p_2$ and $p_3$ yields $p_0$, 
see~(\ref{eq:PMFexample2H0}), (\ref{eq:PMFexample2_2}) and~(\ref{eq:PMFexample2_3}).
Therefore, in the example addressed in the bottom panel of Fig.~\ref{fig:simulPIA01}, the balancing effect described in Sec.~\ref{sec:dacit} cannot occur. In these circumstances,
extending the diffusion mechanism also to agents belonging to different clusters provides additional decision capability. In other words, the larger is $\delta$, the better is the agents' performance, so that choosing $\delta=\infty$, i.e., $\widehat \bcE_k(i)=\cI_k$, would be the best option. For $\delta=\infty$ we only show the performance of agent 1, not to crowd the figure, see the curve in red. 
In summary, if one can exclude the occurrence of the balancing problems described in Sec.~\ref{sec:dacit}, 
then the network design for partially-informed agents simplifies to~(\ref{eq:reold}), with constant combination matrix $A$ satisfying~(\ref{eq:akl}).

As done in the previous subsection for the case of informed agents, we now exploit the results of 
Theorem~\ref{th} and Corollary~\ref{cor} of Sec.~\ref{sec:approx} to provide approximate bounds for the error probability of agent~$k$. 
In particular, exploiting~(\ref{eq:on1}), by Monte Carlo runs we first estimate the value of $\gamma$ appearing in~(\ref{eq:Hoeffding2}) that ensures a type I error probability of $10^{-1}$, 
and then we estimate the correspondent type II error probability. This gives the lower bound.
Exploiting~(\ref{eq:on2}), the same procedure yields the upper bound.
The results are shown in Fig.~\ref{fig:simulboundsPIA}, which refers to the same example of Fig.~\ref{fig:simulPIA01}. 
The error probability for agents~1 and~4, already shown in Fig.~\ref{fig:simulPIA01}, is depicted along with the approximate lower and upper bounds, and an approximation (circles) obtained by
selecting~$\beta_k(A)$ in~(\ref{eq:beta}) in an ad-hoc manner.
Results similar to those shown in Fig.~\ref{fig:simulboundsPIA} are obtained for other agents, with the exception of agents surrounded by many non-effective neighbors
relative to the effective ones (e.g., agents 7, 13, 19), for which the lower bound might be violated.
As in the case of informed agents, our performance prediction is of limited utility for agents lying on cluster borders.

\section{Conclusion}
\label{sec:concl}
In this paper, two multi-task decision problems are considered: one in which agents know the possible states of nature (informed agents) and their task is to 
decide among these time/space-varying multiple possibilities, and another in which agents know the statistics of their observations under a ``normal'' state of nature, and are tasked to detect unknown
time/space-varying deviations from this normal state (partially informed agents). In both cases the decision problem must be resolved without knowledge of the clusters that the agents belong to.
Diffusion mechanisms are designed as modification of the ATC rule, using two continuously updated status components, driven by the locally observed stream of data
combined with data delivered by nearby agents.

Computer simulations show good decision performance at steady-state, even under challenging situations, confirming that the 
clustering mechanism works properly. A theorem is proved that yields a statistical characterization of the agents' status at steady-state, under the simplifying assumption that 
clustering is made without errors. This status characterization is exploited to derive approximate upper and lower bounds for the decision performance.

Because of the underlying assumption of perfect clustering, the derived bounds may be violated in special situations of agents for which the fraction of non-effective neighbors is large. Therefore, our performance prediction may be poor for agents located on cluster borders, which requires further studies. Another future line of research is to develop closer approximations of the system performance,
starting from the status characterization provided in this paper. Insights in this direction have been provided.

\begin{appendix}
\section*{Proof of ThEorem 1} \label{app:proof1}
The Kram\'er-Wold device~\cite[Th. 29.4]{billingsley-book} states that an $R$-dimensional zero-mean random vector $\bq(i)-\E\bq(i)$ converges in distribution to a random vector with distribution $\cN_R(0,\Sigma)$ (when $i\to\infty$) if, and only if, the scalar $u^T [\bq(i)-\E\bq(i)]$ converges in distribution to $\cN(0,u^T\Sigma u)$, for all vectors~$u\in \Re^R$. 
This allows us to generalize the central limit theorem (CLT) for triangular arrays of scalar random variables, see e.g.,~\cite[Th. 27.2]{billingsley-book2},~\cite[Th. 1.15]{shao},
to triangular arrays made of random vectors. However, to apply these results to our case,  
$\lim_{i\to\infty}$ should be replaced with the double limit $\lim_{\mu\to0}\lim_{i\to\infty}$, which
can be addressed as done in~\cite[App.~B]{MaranoSayedIT19}. 
This way, we obtain the version of the Lindeberg-Feller CLT for arrays of vectors given in~\cite[Prop.\ 2.27]{vandervaart} under a double limit formulation, which reads as follows.
With reference to Theorem~\ref{th}, fix $k$, $i$, and $j$, and note that each entry of the vector $\bq_k(i,j)$ in~(\ref{eq:array}) has finite variance because $p_h(a)>0$, $\forall a \in \cA$,
implies $|\bd_k^{(r)}(\cdot)|^2 \le N$ for all $r,k$, and some $N>0$.
Note also that the elements on each row of the array~(\ref{eq:array}) are mutually independent.
Let $\mathbb{COV}(\cdot)$ denote the covariance matrix of a random vector, and let $\mathbb{I}(\cdot)$ be the indicator function.
If, for some covariance matrix $\Sigma$ and every $\epsilon>0$,
\begin{align} 
& \lim_{\mu\to 0} \lim_{i\to \infty} \sum_{j=1}^i \mathbb{COV}(\bq_k(i,j)) = \Sigma, \label{eq:if1}\\
& \lim_{\mu\to 0} \lim_{i\to \infty} \sum_{j=1}^i \E \left [ \| \bq_k(i,j) \|^2 \,  \mathbb{I} \big (\|\bq_k(i,j) \| >\epsilon \big )\right ] =0, \label{eq:if2}
\end{align}
then $\sum_{j=1}^i \big [ \bq_k(i,j)- \E\bq_k(i,j) \big ]$ converges in distribution to a zero-mean $R$-vector with covariance matrix $\Sigma$, 
for $i\to \infty$ followed by $\mu\to 0$.

Consider the expression in~(\ref{eq:if1}). 
By exploiting the independence of the observations $\bx_k(i)$ for different values of~$k$, 
simple calculations show that the covariance of the two variables $\bt_k^{(r_1)}(i,j)$ and $\bt_k^{(r_2)}(i,j)$ appearing in~(\ref{eq:ta12}), is
\begin{align}
\mu (1-\mu)^{2j-2} \sum_{\ell=1}^S  b_{k\ell}^2(j) \, [\Lambda_k]_{mn}, \label{eq:anchetu}
\end{align} 
where~(\ref{eq:anchetu}) exploits the IID property of observations from the same cluster.
By summing~(\ref{eq:anchetu}) for $j$ ranging from 1 to $i$, and taking the limits $\lim_{\mu \to 0} \lim_{i \to \infty}$, yields~$\Sigma=\beta_k(A) \, \Lambda_k$, see~(\ref{eq:lambda})-(\ref{eq:beta}).

Consider next Lindeberg condition~(\ref{eq:if2}), and fix $\delta>0$. By omitting the indexes $(i,j)$ and the subscript $k$ for notational simplicity, we have
{\small \[ 
\E \left [ \| \bq \|^2 \,  \mathbb{I} \big (\|\bq \| >\epsilon \big )\right ] \le \E \left [ \frac{\| \bq \|^{2+\delta}}{\epsilon^\delta} \,  \mathbb{I} \big (\|\bq \| >\epsilon \big )\right ]
 \le \frac{\E  \left [ \| \bq \|^{2+\delta} \right ]}{\epsilon^\delta},
 \]}%
which shows that
$\lim_{\mu \to 0}\lim_{i \to \infty}\sum_{j=1}^i\E  \left [ \| \bq_k(i,j) \|^{2+\delta} \right ]=0$, known as  Lyapunov condition~\cite{billingsley-book2}, 
implies condition~(\ref{eq:if2}). 
Now, from~(\ref{eq:ta1})-(\ref{eq:vect}), 
{\small\begin{subequations}\begin{align} 
&\E  \left [ \| \bq_k(i,j) \|^{2+\delta} \right ] = \mu^{1+\frac \delta 2} (1-\mu)^{(j-1)(2+\delta)} \nonumber \\
& \qquad\qquad \times \E \Bigg \{  \sum_{r=1}^R \Bigg [\sum_{\ell=1}^S b_{k\ell}(j)  \bd_\ell^{(r)}(i-j+1) \Bigg ]^2    \hspace*{-1pt}\Bigg \}^{1+\frac \delta 2} \label{eq:long1} \\
& \le \mu^{1+\frac \delta 2} (1-\mu)^{(j-1)(2+\delta)} \nonumber \\
& \qquad \quad \times \E \Bigg \{ \sum_{r=1}^R \Bigg [\sum_{\ell=1}^S  b_{k\ell}^2(j) \sum_{\ell=1}^S \Big( \bd_\ell^{(r)} (i-j+1)\Big )^2  \Bigg ] \Bigg \}^{1+ \frac \delta 2} 
\hspace*{-25pt}\label{eq:long2} \\
& \le \mu^{1+\frac \delta 2} (1-\mu)^{(j-1)(2+\delta)} ( RSN )^{1+ \frac \delta 2}, \label{eq:long3}
\end{align} \label{eq:long}%
\end{subequations}}%
where (\ref{eq:long2}) follows by Cauchy-Schwarz inequality, and (\ref{eq:long3}) from $\sum_{\ell=1}^S b_{k\ell}^2(j) \le 1$, and  $|\bd_\ell^{(r)}(\cdot)|^2 \le N$. 
Inequality~(\ref{eq:long}) shows that condition~(\ref{eq:if2})
is verified because
{\small \begin{align}
\lim_{\mu \to 0} \sum_{j=1}^\infty \mu^{1+\frac \delta 2} (1-\mu)^{(j-1)(2+\delta)} 
= \lim_{\mu \to 0} \frac{\mu^{1+\frac \delta 2}}{1-(1-\mu)^{2+\delta}}=0. \nonumber
\end{align}}%
The bounds in~(\ref{eq:bounds}) follow by recalling that the combination matrix~$A$ is nonnegative and right-stochastic and so are its 
powers~$B(i)=A^i$, yielding $\sum_{\ell=1}^S b_{k\ell}(i)=1$, $\forall i$. The upper bound follows immediately, and the lower bound follows by Cauchy-Schwarz inequality
$(\sum_{\ell=1}^S b_{k\ell}(i))^2 \le S \sum_{\ell=1}^S b_{k\ell}^2(i)$.
 
\end{appendix}



\end{document}